%% file: main.tex
\documentclass{article}

\usepackage{graphicx}
\usepackage{amsmath}
\usepackage{amsthm}
\usepackage{natbib}
\usepackage{amsfonts}
\usepackage{hyperref}
\usepackage{authblk}
\usepackage[verbose=true]{geometry}
\usepackage{float}
\usepackage{enumerate}
\usepackage[inline]{enumitem}
\usepackage{url}
\usepackage{bbm}
\usepackage{bm}
\usepackage{array}
\usepackage{tikz}

\newgeometry{
    textheight=9in,
    textwidth=5.5in,
    top=1in,
    headheight=12pt,
    headsep=25pt,
    footskip=30pt
}

\theoremstyle{plain}
\newtheorem{theorem}{Theorem}

\newcommand{\E}{E}
\newcommand{\Cov}{\mathrm{cov}}
\newcommand{\Var}{\mathrm{var}}
\newcommand{\indep}{\perp\!\!\!\perp}

\graphicspath{{plots/}}

\DeclareMathOperator*{\argmin}{arg\,min}

\DeclareMathOperator*{\logit}{logit}

\newcounter{assumption_list}

\newcommand{\mytitle}{Variable importance measures for heterogeneous treatment effects}

\title{\mytitle}
\date{\today}
\author[1]{Oliver J.~Hines}
\author[2]{Karla Diaz-Ordaz}
\author[3]{Stijn Vansteelandt}
\affil[1]{Columbia University, New York, NY, USA}
\affil[2]{University College London, London, U.K.}
\affil[3]{Ghent University, Ghent, Belgium}

\begin{document}

\maketitle

\begin{abstract}
Motivated by applications in precision medicine and treatment effect heterogeneity, recent research has focused on estimating conditional average treatment effects (CATEs) using machine learning (ML). CATE estimates may represent complicated functions that provide little insight into the key drivers of heterogeneity. Therefore, we introduce nonparametric treatment effect variable importance measures (TE-VIMs), based on the mean-squared error (MSE) in predicting the individual treatment effect. More precisely, TE-VIMs represent the increase in MSE when variables are removed from the CATE conditioning set. We derive efficient TE-VIM estimators which can be used with any CATE estimation strategy and are amenable to ML estimation. We propose several strategies to calculate these VIMs (e.g. leave-one out, or keep-one in), using popular meta-learners for the CATE. We study the finite sample performance through a simulation study and illustrate their application using clinical trial data.
\end{abstract}

Keywords: Causal inference; Conditional effects; Data-adaptive estimation; Effect modification.

\section{Introduction}
\label{sect:intro}

In the medical and social sciences there is a longstanding interest in quantifying the heterogeneity in the effect of a treatment or intervention on a population. 
Understanding such heterogeneity is essential for informing scientific research and optimizing treatment decisions.
Attention focused initially on subgroup analyses, which identify population subgroups that benefit most/least from treatment, to be evaluated further in potential future studies \citep{Rothwell2005, Slamon2001}.
Typical challenges of subgroup analyses are selecting stratification variables in a systematic way and handling the resulting multiplicity problem. Endeavours to address these were soon followed by methodological developments on personalized medicine in causal inference, pioneered by \cite{Murphy2003}, with the primary focus being on policy learning; i.e., determining the %optimal
treatment policy that minimizes some measure of population risk
\citep{VanderLaan2014,Athey2021}.

Recently, attention has partially shifted towards learning the conditional average treatment effect (CATE) $\tau(\bm{x}) \equiv \E(Y^1-Y^0|\bm{X}=\bm{x})$, where $Y^a$ is the outcome observed if treatment $A$ were set to $ a\in \{0,1\}$, and $\bm{X} \in \mathbb{R}^p$ are pre-treatment covariates \citep{Athey2016,Wager2018,Kunzel2019,Kennedy2020}. The CATE provides insight into the magnitude of the treatment effect for each individual and the optimal dynamic treatment rule (OTR), obtained from the sign of the CATE \citep{vanderweele2019selecting}.

These foregoing developments are important, but leave unanswered a key question: what are the key drivers of treatment effect heterogeneity? Answers of this question may inform about treatment mechanism, suggest future therapies, help compare clinical trial populations, or help quantify systematic treatment biases (e.g. due to race or socio-economic status). 
One pioneering proposal for CATE variable attribution is to extend random forest variable importance measures (VIMs) through the `causal forest' estimator of the CATE \citep{Athey2019,Wager2018}. The resulting VIMs rely on the tree architecture of causal forests, and may inherently assign greater importance to continuous variables, or categorical variables with many categories \citep{Gromping2009}.
VIMs based on specific modeling strategies (decisions trees, linear regression, etc.) are referred to as `algorithmic' \citep{Williamson2020}. Whilst algorithmic-VIMs can provide useful insights, there remains a need for model-agnostic alternatives, especially when the chosen CATE estimating strategy does not have a well-established algorithmic-VIM, but also in order to compare VIMs following different CATE estimation procedures.
Therefore, we propose treatment effect VIMs (TE-VIMs) that are nonparametric summary statistics, which measure the importance of variable subsets in predicting the individual treatment effect (ITE) $Y^1 - Y^0$. TE-VIMs are relatively easy to communicate to researchers already familiar with traditional goodness-of-fit methods such as ANOVA, and, unlike algorithmic-VIMs, allow researchers to compare heterogeneous treatment effect insights across different CATE learning algorithms.

More precisely, we consider the mean-squared-error $L\{f\} \equiv \E[\{Y^1 - Y^0 - f(\bm{X})\}^2]$, which for arbitrary $f:\mathbb{R}^p \mapsto \mathbb{R}$ is not identified without strong assumptions on the joint distribution of $(Y^1, Y^0)$ \citep{Levy2021,Ding2016}.
%Heckman1997
One key insight is that $L\{f\} = \E[\{\tau(\bm{X}) - f(\bm{X})\}^2] + \E\{\Var(Y^1-Y^0|\bm{X})\}$ comprises a first term that is identified under standard causal assumptions and another that does not depend on $f$. Exploiting this decomposition, we define the TE-VIM
$\Theta_s \equiv  L\{\tau_s\} - L\{\tau\}$, where $\tau_s(\bm{x}) \equiv \E(Y^1-Y^0|\bm{X}_{-s}=\bm{x}_{-s})$ is the CATE conditional on $\bm{X}_{-s}$, and $\bm{u}_{-s}$ denotes the vector of all the components of $\bm{u}$ with index not in $s \subseteq \{1,...,p\}$. Note that $\tau_s(\bm{x})$ only depends on $\bm{x}_{-s}$, but we write it as a function of $\bm{x}$ to simplify notation. We interpret $\Theta_s = \Var\{\tau(\bm{X})\} - \Var\{\tau_s(\bm{X})\}$ in terms of the variance $\Var\{\tau(\bm{X})\}$ of the treatment effect (VTE), a global measure of treatment effect heterogeneity due to \cite{Levy2021}, see Supplement \ref{supp:representations}.
Specifically, $\Theta_s \geq 0$ represents the increase in CATE variance when variables in $s$ are excluded from the conditioning set. Thus, $\Theta_s$ quantifies the treatment effect heterogeneity explained by $\bm{X}_s$, beyond that already explained by $\bm{X}_{-s}$, where $\bm{u}_s$ denotes the vector of all components of $\bm{u}$ with index in $s$.

Moreover, TE-VIMs connect to regression-VIMs  \citep{Williamson2021,Zhang2020}, also called `leave-out covariates' \citep{Lei2018, Verdinelli2021}, and the nonparametric VIM framework of \cite{Williamson2021_general}. The latter framework covers VIMs that represent differences in value (negative risk) functions.
Our work represents a step towards applying this framework in more complicated settings, such as in causal inference where identification may be a concern.

In Section \ref{sec_methods} we derive efficient TE-VIM estimators and motivate the DR-learner of the CATE  by interpreting our estimators in terms of pseudo-outcomes \citep{Kennedy2020}. Results on simulated data are presented in Section \ref{sect:simulation} and in Section \ref{sect:example} we use TE-VIMs to identify drivers of treatment effect heterogeneity in a clinical trial setting. In Section \ref{sect:extend} we compare TE-VIMs with existing OTR-VIMs and outline extensions to continuous treatments.

\section{Methodology}
\label{sec_methods}

\subsection{Motivating the estimand}

We consider $n$ i.i.d. observations $(\bm{z}_1,...,\bm{z}_n)$ of a random variable $\bm{Z}=(Y,A,\bm{X}) \sim P_0$ distributed according to an unknown distribution $P_0\in\mathcal{M}$ and consisting of an `outcome' $Y\in \mathbb{R}$, an `exposure' or `treatment' $A\in\{0,1\}$, and covariates $\bm{X} \in \mathbb{R}^p$.
In a slight abuse of notation, we let $p$ denote the index set $\{1,...,p\}$ so that $\tau_p$ is the average treatment effect (ATE) and $\Theta_{p}$ is the VTE, which we assume to be non-zero. Assuming consistency ($A=a \implies Y=Y^a$), conditional exchangeability ($Y^a \indep A|\bm{X}$ for $a=0,1$), and positivity ($0<\pi(\bm{X})<1$ almost surely), the CATE is identified by $\tau(\bm{x}) =\mu(1,\bm{x})-\mu(0,\bm{x})$, where $\mu(a,\bm{x})\equiv E(Y|A=a,\bm{X}=\bm{x})$ and $\pi(\bm{x})\equiv \E(A|\bm{X}=\bm{x})$ is the `propensity score'. We let $||f(\bm{Z})|| \equiv \E[\{f(\bm{Z})\}^2]^{1/2}$ denote the $L_2(P_0)$ norm of an arbitrary function $f$, and when $f$ is estimated from data, the expectation is taken only over the random inputs $\bm{Z}$. Assuming that $||\tau(\bm{X})|| < \infty$, then the TE-VIM $\Theta_{s} = ||\tau(\bm{X}) - \tau_s(\bm{X})||^2  <\infty$ is also finite. By construction, $s^\prime \subseteq s \subseteq p$ implies $\Theta_{s^\prime} \leq \Theta_s \leq \Theta_{p}$, i.e. the set $s^\prime$ cannot be more important than $s$. Generally, the covariates used to define the CATE need not be the same as the covariates $\bm{X}$ required for conditional exchangeability to hold. E.g. one could consider the target estimand $L\{\tau_{s^\prime}\} - L\{\tau_s\} = \Theta_{s^\prime} - \Theta_{s} $, which quantifies the importance of $s^\prime$ with $s$ treated as the full covariate set. This extension follows from results for $\Theta_{s}$, which we study in the current work.

The TE-VIM $\Theta_s$ is analogous to the regression-VIM $\Omega_s \equiv ||Y-\mu_s(\bm{X})||^2 - ||Y-\mu(\bm{X})||^2$ due to \cite{Williamson2021}, which replaces the ITE $Y^1-Y^0$ with $Y$, and hence $\tau(\bm{x})$ with $\mu(\bm{x})\equiv\E(Y|\bm{X}=\bm{x})$ and $\tau_s(\bm{x})$ with $\mu_s(\bm{x})\equiv\E(Y|\bm{X}_{-s}=\bm{x}_{-s})$. The two proposals differ, however, in how the VIM is scaled with \cite{Williamson2021} defining the scaled regression-VIM $\Omega_s/\Var(Y)$ in analogy with the familiar $R^2$ statistic. Since $\Var(Y^1-Y^0)$ is not easily identified, we instead propose scaled TE-VIMs as $\Psi_s \equiv \Theta_s/\Theta_p = 1 - \Var\{\tau_s(\bm{X})\} / \Theta_p$. Like an $R^2$ statistic, we interpret $\Psi_s \in [0,1]$ as the proportion of treatment effect heterogeneity explained by $\bm{X}_{-s}$ compared with $\bm{X}$. For instance, under the linear model $\E(Y^a|\bm{X}=\bm{x}) = \mu(a, \bm{x}) = \beta(\bm{x}) + a \tau(\bm{x})$, where $\beta(\bm{x})$ and $\tau(\bm{x})$ are both linear in $\bm{x}$, $\Psi_s$ is the limiting $R^2$ value obtained from a linear regression of the effect modifier $\tau(\bm{X})$ on $\bm{X}_{-s}$. Moreover, $\Psi_s$ and $\Omega_s/\Var(Y)$ are both invariant to linear transformations of the outcome and invertible component-wise transformations of $\bm{X}$, see Supplement \ref{supp:invariance} for details.
In practice, VIM scaling makes little difference when using $\Psi_s$ and $\Psi_{s^\prime}$ to compare the relative importance of $s$ and $s^\prime$.
Instead, the main decision for investigators is which covariate sets should be compared, and we identify the following modes of operation in this regard:
\begin{enumerate}
    \item \textbf{Leave-one-out (LOO)}: The set $s$ contains a single covariate of interest. This mode may `under represent' importance when covariates are highly correlated.
    \item \textbf{Keep-one-in (KOI)}: The set $s$ contains all but a single covariate of interest. This mode may `over represent' importance when covariates are highly correlated, and is less sensitive to multi-covariate interactions.
    \item \textbf{Shapley values}: TE-VIMs for all possible $(2^p)$ covariate combinations are considered and aggregated in a game theoretic manner \citep{Owen2017,Williamson2020}. This is a theoretically appealing compromise between LOO and KOI, but may be computationally impractical for even modest $p$ and render clinical interpretation more subtle. A definition of TE-VIM Shapley values is given in Supplement \ref{supp:shapley}.
    \item \textbf{Covariate grouping}: Domain specific knowledge is used to group covariates, simplifying the above modes by considering covariates block-wise (e.g. comparing biological vs. non-biological factors).
\end{enumerate}

Note that we use `under' and `over' represent in a relative sense, since the ground truth variable importance depends on the importance definition used. See e.g. \cite{Hama2022, Verdinelli2023} for recent discussions on this topic and \cite{Tang2023} for variable selection proposals. In Section \ref{sect:example} we demonstrate the LOO and KOI modes through an applied example.

\subsection{CATE estimation}
\label{sect_cates}

Estimation of TE-VIMs will rely on initial CATE estimates, obtained via flexible machine learning based methods, which we review first. CATE estimation is challenging since common machine learning algorithms (random forests, neural networks, boosting etc.) are designed for mean outcome regression, e.g. by minimizing the mean squared error loss. CATE estimation strategies therefore either modify existing machine learning methods to target CATEs, e.g. \cite{Athey2019} modify the random forest algorithm for CATE estimation. Alternatively, `metalearning' strategies decompose CATE estimation into a sequence of sub-regression problems, which can be solved using off-the-shelf machine learning algorithms \citep{Kunzel2019,Kennedy2020}.

In the current work we focus on two metalearning algorithms which, following the naming convention of \cite{Kunzel2019} and \cite{Kennedy2020}, we refer to as the T-learner and the DR-learner. The T-learner is based on the decomposition $\tau(\bm{x})=\mu(1,\bm{x})-\mu(0,\bm{x})$, and estimates the CATE by $\hat{\tau}^{(T)}(\bm{x})\equiv \hat{\mu}(1,\bm{x})-\hat{\mu}(0,\bm{x})$, where $\hat{\mu}(a,\bm{x})$ represents an estimate of $\mu(a,\bm{x})$ obtained by a regression of $Y$ on $X$ using observations where $A=a$.
The T-learner, however is problematic for two main reasons. Firstly, whilst regularization methods can be used to control the smoothness of $\hat{\mu}(a,\bm{x})$, the same is not true of $\hat{\tau}^{(T)}(\bm{x})$ which may be erratic. Slow convergence rates affecting $\hat{\mu}(a,\bm{x})$ may therefore propagate into $\hat{\tau}^{(T)}(\bm{x})$. 
Secondly, $\hat{\mu}(1,\bm{x})$ is chosen to make an optimal bias-variance trade-off over the covariate distribution of the treated population. Likewise, $\hat{\mu}(0,\bm{x})$ is chosen to make an optimal bias-variance trade-off over the covariate distribution of the untreated population. When there is poor overlap between the treated and untreated subgroups, then $\hat{\tau}^{(T)}(\bm{x})$ may fail to deliver an optimal bias-variance trade-off over the population covariate distribution, making the T-learner potentially poorly targeted towards CATE estimation.

The DR-learner \citep{Kennedy2020,Luedtke2016,VanderLaan2013} is an alternative metalearning algorithm based on the decomposition $\tau(\bm{x})=\E\{\varphi(\bm{Z})|\bm{X}=\bm{x}\}$ where, for $\bm{z} = (y,a,\bm{x})$ 
\begin{align}
\varphi(\bm{z}) \equiv \{y-\mu(a,\bm{x})\}\frac{a-\pi(\bm{x})}{\pi(\bm{x})\{1-\pi(\bm{x})\}} + \mu(1,\bm{x}) - \mu(0,\bm{x}) \label{po_definition}.
\end{align}
is called the `pseudo outcome', or the augmented inverse propensity weighted score \citep{Robins1994}, and acts like the ITE in expectation. The DR-learner first estimates $\mu(a,\bm{x})$ and $\pi(\bm{x})$ to obtain the pseudo-outcome estimator $\hat{\varphi}(\bm{Z})$, where $\mu(a,\bm{x})$ and $\pi(\bm{x})$ in \eqref{po_definition} are replaced with estimates $\hat{\mu}(a,\bm{x})$ and $\hat{\pi}(\bm{x})$.

In a second step, the estimated pseudo-outcome, $\hat{\varphi}(\bm{Z})$ is regressed on covariates $\bm{X}$ to obtain $\hat{\tau}^{(DR)}(\bm{x})$. A sample splitting scheme is also recommended, whereby the regression steps to obtain $\hat{\mu}(a,\bm{x})$, $\hat{\pi}(\bm{x})$, and $\hat{\tau}^{(DR)}(\bm{x})$ are performed on three independent samples.

The DR-learner alleviates the issues related to the T-learner since the complexity of $\hat{\tau}^{(DR)}(\bm{x})$ can be controlled by regularizing the regression in the final stage of the procedure, mitigating concerns regarding the smoothness of the T-learner. With regard to consistency, taking expectation over the random inputs $\bm{Z}$, the square of $\E\{\hat{\varphi}(\bm{Z})|\bm{X}=\bm{x}\}-\tau(\bm{x})$ is bounded above by the product of the squared estimation errors of the propensity score and regression estimators (up to constant scaling). In practice, this means that the regression of $\hat{\varphi}(\bm{Z})$ on $\bm{X}$ mimics the oracle regression of $\varphi(\bm{Z})$ on $\bm{X}$ provided that
\begin{enumerate}[label=(A\arabic*)]
\item $||\{\pi(\bm{X})-\hat{\pi}(\bm{X})\}\{\mu(a,\bm{X})-\hat{\mu}(a,\bm{X})\}|| = o_P(n^{-1/2})$ for $a = 0, 1$.

\setcounter{assumption_list}{\value{enumi}}
\end{enumerate}
and a suitable cross-fitting procedure is used. (A1) implies that one can trade-off accuracy in the outcome and propensity score estimators, a property known as double robustness, hence the name `DR-learner'. Note that the $n^{-1/2}$ rate in (A1) differs from the `stability' condition provided for the DR-learner by \citet[Proposition 1]{Kennedy2020}, as it reflects what will be required by our TE-VIM estimators in Section \ref{sect:tevim_estimation}.

Estimation of the CATE $\tau_s(\bm{x})$ is complicated by the fact that one cannot assume that $Y\indep A|\bm{X}_{-s}$ for an arbitrary subset of covariates $s$, a problem that is sometimes referred to as `runtime confounding' \citep{Coston2020}. The DR-learner readily accommodates runtime confounding through the identity $\tau_s(\bm{x})=\E\{\varphi(\bm{Z})|\bm{X}_{-s}=\bm{x}_{-s}\}$. This identity implies that one may estimate $\tau_s(\bm{x})$ by regressing $\hat{\varphi}(\bm{Z})$ on $\bm{X}_{-s}$, i.e. modifying the final regression step of the DR-learner.

We recommend a metalearner for $\tau_s(\bm{x})$ based on the identity $\tau_s(\bm{x})=\E\{\tau(\bm{X})|\bm{X}_{-s}=\bm{x}_{-s}\}$. Specifically, we propose estimating $\tau_s(\bm{x})$ by regressing initial CATE estimates, $\hat{\tau}(\bm{X})$ on $\bm{X}_{-s}$ via a machine learning algorithm. Like the DR-learner, one can regularize the resulting CATE estimator $\hat{\tau}_s(\bm{x})$. We advocate this approach, since it usually results in estimates of $\tau_s(\bm{x})$ which are compatible with those of $\tau(\bm{x})$, in the sense of respecting the fact that the two conditional means are related. For instance, we expect that $\E\{\varphi(\bm{Z})\} = \E\{\tau(\bm{X})\} = \E\{\tau_s(\bm{X})\}$, but these equalities are generally violated for the corresponding estimated quantities $\E\{ \hat{\varphi}(\bm{Z})\}$, $\E\{ \hat{\tau}(\bm{X})\}$, and $\E\{ \hat{\tau}_s(\bm{X})\}$.

\subsection{TE-VIM estimation}
\label{sect:tevim_estimation}

\subsubsection{Estimation of $\Theta_s$}

We consider estimators based on the efficient influence curve (IC) of $\Theta_s$ under the nonparametric model. Briefly, ICs are mean zero functions that characterize the sensitivity of so-called pathwise differentiable estimands to small changes in the data generating law. Thus, ICs are useful for constructing efficient estimators and determining their asymptotic distribution, see e.g. \cite{Hines2021} for an introduction. 

TE-VIMs  fall under the VIM framework of \cite{Williamson2021_general}, for which generic IC results are available. These results cannot be directly applied, however, since the risk $L\{f\}$ is not identified.
Instead, we consider that $\Theta_s = L_U\{\tau_s\} - L_U\{\tau\}$ where
\begin{align*}
    L_U\{f\} = \E[\{U + \tau(\bm{X})- f(\bm{X})\}^2] = \E[\{\tau(\bm{X})- f(\bm{X})\}^2] + \E(U^2)
\end{align*}
and $U$ is a random variable such that $\E(U|\bm{X}) = 0$. Note that $U = Y^1 - Y^0 - \tau(\bm{X})$ recovers the unidentified risk $L\{f\}$ and $U = \varphi(\bm{Z}) - \tau(\bm{X})$ recovers the DR-learner population risk. To simplify derivations we consider the risk $L^*\{f\}$ obtained by setting $U=0$. Theorem 3 of  \cite{Williamson2021_general} states that, for this risk, there is no price to pay to estimate its minimizer $\tau_s(\bm{x})$, insofar as the IC for $L^*\{\tau_s\}$ that is derived when $\tau_s(\bm{x})$ is known is the same as that derived when $\tau_s(\bm{x})$ is unknown. In Supplement \ref{ic_deriv_append} we use point mass contamination to show that, for known $f$, the IC of $L^*\{f\}$ at a single observation $\bm{z}$ is $\{\varphi(\bm{z})-f(\bm{x})\}^2 - \{\varphi(\bm{z})-\tau(\bm{x})\}^2  - L^*\{f\}$. Hence, applying the aforementioned Theorem, the IC of $\Theta_s$ is
\begin{align}
\phi_s(\bm{z}) &\equiv \{\varphi(\bm{z})-\tau_s(\bm{x})\}^2 - \{\varphi(\bm{z})-\tau(\bm{x})\}^2  - \Theta_{s}. \label{IC:theta}
\end{align}

The interpretation of $\varphi(\bm{Z})$ as a pseudo-outcome, which plays the role of the unobserved ITE $Y^1-Y^0$, holds in the present context. To see why, we compare \eqref{IC:theta} to the IC of $\Omega_s$ by \cite{Williamson2021}, $\{y-\mu_s(\bm{x})\}^2 - \{y-\mu(\bm{x})\}^2  - \Omega_s$, which is of the same form as \eqref{IC:theta}, but with the outcome $y$ replacing the pseudo-outcome $\varphi(\bm{z})$.

The IC in \eqref{IC:theta} may be used to construct efficient estimating equation estimators of $\Theta_s$ by setting (an estimate of) the sample mean IC to zero. This strategy is equivalent to the so-called one-step correction outlined in Supplement \ref{Appen_assymp}. We thus obtain the estimator
\begin{align}
\hat{\Theta}_{s} &\equiv n^{-1}\sum_{i=1}^n \left[\{\hat{\varphi}(\bm{z}_i)-\hat{\tau}_s(\bm{x}_i)\}^2 - \{\hat{\varphi}(\bm{z}_i)-\hat{\tau}(\bm{x}_i)\}^2\right], \label{Est:theta}
\end{align}
where superscript hat denotes consistent estimators. In practice, we recommend a cross-fitting procedure of the type described in Algorithms SS-A and SS-B, to obtain the fitted models and evaluate the estimators using a single sample \citep{Chernozhukov2018,van_der_laan_cross-validated_2011}. We discuss the reasons for sample splitting with reference to Theorem \ref{asym_theorem}, which gives conditions under which $\hat{\Theta}_s$ is regular asymptotically linear (RAL).

\begin{theorem}
\label{asym_theorem}

Assume that there exist constants $\epsilon,K,\delta \in(0,\infty)$ such that almost surely $\hat{\pi}(\bm{X}) \in (\epsilon, 1-\epsilon), \Var\{\varphi(\bm{Z})|\bm{X}\} < K$, and $||\hat{\tau}(\bm{X}) - \hat{\tau}_s(\bm{X})|| < \delta$. Suppose also that at least one of the following two conditions hold:
\begin{enumerate}[label=\arabic*.]
\item \textbf{Sample splitting}: $\hat{\pi}(\bm{x}), \hat{\mu}(\bm{x}), \hat{\tau}(\bm{x})$, and $\hat{\tau}_s(\bm{x})$ are obtained from a sample independent of the one used to construct $\hat{\Theta}_s$.
\item \textbf{Donsker condition}: The quantities $\{\varphi(\bm{Z}) - \hat{\tau}(\bm{X})\}^2$, $\{\varphi(\bm{Z}) - \hat{\tau}_s(\bm{X})\}^2$, and $\{\hat{\tau}(\bm{X}) - \hat{\tau}_s(\bm{X})\}\hat{\varphi}(\bm{Z})$ fall within a $P_0$-Donsker class with probability approaching $1$.
\end{enumerate}
Finally assume (A1) holds, and (A2) that $||\tau(\bm{X}) -\hat{\tau}(\bm{X})||$ and $||\tau_s(\bm{X})-\hat{\tau}_s(\bm{X})||$ are both $o_P(n^{-1/4})$. Then $\hat{\Theta}_{s}$ is asymptotically linear with IC $\phi_s(\bm{Z})$, and hence $\hat{\Theta}_{s}$ converges to $\Theta_{s}$ in probability, and for $\Theta_{s} >0$ then $n^{1/2}(\hat{\Theta}_{s}-\Theta_{s})$ converges in distribution to a mean-zero normal random variable with variance $||\phi_s(\bm{Z})||^2$.
\end{theorem}

Assumptions (A1-A2) both require nuisance function estimators to converge at sufficiently fast rates. The requirement for $n^{1/4}$ rate convergence in (A2) is standard in the recent VIM framework of \cite{Williamson2021_general}, whilst (A1) is additionally required to control for errors in estimating the pseudo-outcomes.

Together, (A1-A2) suggest that the DR-learner may be preferred over the T-learner due to its robustness. In particular, the T-learner of the CATE satisfies $||\tau(\bm{X}) - \hat{\tau}^{(T)}(\bm{X})|| = o_P(n^{-1/4})$, provided that $||\mu(a,\bm{X})-\hat{\mu}(a,\bm{X})|| = o_P(n^{-\alpha})$, with $\alpha \geq 1/4$ for $a=0,1$. (A1) then implies that $||\pi(\bm{X})-\hat{\pi}(\bm{X})||$ must be at least $o_P(n^{-1/2 + \alpha})$. I.e. the propensity score estimator may converge at a slower rate, if the outcome estimator converges at a faster rate, but the converse is not true since fast convergence of the outcome estimator is needed to assure sufficiently fast convergence of the T-learner. This is unsatisfying for example in clinical trial settings, where the exposure is randomized and propensity scores are known.

The DR-learner, however, satisfies $||\tau(\bm{X}) - \hat{\tau}^{(DR)}(\bm{X})|| =o_P(n^{-1/4})$ when (A1) holds and $||\E\{\hat{\varphi}(\bm{Z})|\bm{X}\}-\hat{\tau}^{(DR)}(\bm{X})||= o_P(n^{-1/4})$, i.e. when the final DR-learning regression estimator is consistent at $n^{-1/4}$ rate. Applying the same reasoning as before, (A1) implies that $||\mu(A,\bm{X})-\hat{\mu}(A,\bm{X})||$ can be $o_P(n^{-\alpha})$, if $||\pi(\bm{X})-\hat{\pi}(\bm{X})||$ is $o_P(n^{-1/2 + \alpha})$, for any $\alpha \in(0,1/2)$. In other words, the outcome estimator is allowed to converge at a slower rate, provided the propensity score estimator converges at a faster rate and vice-versa, which marks an improvement over the T-learner, at the expense of an additional requirement on the final DR-learning step. The requirement on the DR-learning step, however, will likely be weaker than that on outcome estimator, since $\tau(\bm{x})$ is likely smoother than $\mu(a,\bm{x})$, e.g. when the CATE depends only on a subset of $\bm{X}$.

The Donsker condition in Theorem \ref{asym_theorem} controls the `empirical process' term in the estimator expansion \citep{Newey2018,Hines2021}. This condition is usually not guaranteed to hold when flexible machine learning methods are used to estimate nuisance functions. Fortunately, cross-fitting using two or more splits of the data offers a way of avoiding Donsker conditions, at the expense of making nuisance functions more computationally expensive to learn \citep{Chernozhukov2018,van_der_laan_cross-validated_2011}.

\subsubsection{Importance testing}
\label{sect:import}

One property shared by $\hat{\Theta}_{s}$ and the analogous $\Omega_s$ estimator \citep{Williamson2021} concerns their behavior under the zero-importance null hypothesis, $H_0: \Theta_{s} = 0$. For TE-VIMs, $H_0$ corresponds to treatment effect homogeneity $\tau(\bm{x})=\tau_s(\bm{x})$, in which case $\phi_s(\bm{Z}) = 0$. This IC degeneracy makes $H_0$ difficult to test, since $\hat{\Theta}_s$ is not necessarily asymptotically normal. For this reason, Theorem \ref{asym_theorem} considers the asymptotic distribution only when $\Theta_{s} > 0$.
One solution to the IC degeneracy problem is to estimate $\Var\{\tau(\bm{X})\}$ and $\Var\{\tau_s(\bm{X})\}$ using efficient estimators in separate samples \citep{Williamson2021_general}. Each estimand has a non-zero IC provided that $\Var\{\tau_s(\bm{X})\}>0$, despite both ICs being identical under $H_0$. Thus both estimators are independent and asymptotically normal, hence their difference (an estimator of $\Theta_s$) is also asymptotically normal even when $\Theta_s=0$. One therefore obtains a valid Wald-type test for $H_0$, at the expense of using an estimator for $\Theta_s$, which is inefficient because the component estimators are estimated using only half of the data. Similarly, one could test the zero-VTE null hypothesis $(\Var\{\tau(\bm{X})\}=0)$ by estimating $\E\{\tau^2(\bm{X})\}$ and $\E\{\tau(\bm{X})\}^2$ using efficient estimators in separate samples and taking their difference. Such approaches are an active area of research and evaluating them in the context of TE-VIMs is beyond the scope of the current work \citep{Hudson2023, Guo2024}. Moreover, the distribution of $\hat{\Theta}_s$ under $H_0$ may depend on higher-order pathwise derivatives of the estimand, which are also an open research area \citep{Carone2018}.

\subsubsection{Estimation of $\Psi_s$}

The scaled TE-VIM $\Psi_s\equiv\Theta_s/\Theta_p$, has IC, $\Phi_s(\bm{z}) \equiv \{\phi_s(\bm{z}) - \Psi_{s}\phi_{p}(\bm{z})\}/\Theta_{p}$, where $\phi_p(\bm{z})$ denotes \eqref{IC:theta} for the index set $s=p$. This IC implies an estimating equations estimator, $\hat{\Psi}_{s} = \hat{\Theta}_{s}/\hat{\Theta}_{p}$, where $\hat{\Theta}_{p}$ is the VTE estimator obtained when $\hat{\tau}_s(\bm{x})$ in \eqref{Est:theta} is replaced with an ATE estimate $\hat{\tau}_p$.

The estimators $\hat{\Theta}_p$ and $\hat{\Psi}_s$ both rely on an ATE estimator $\hat{\tau}_p$. The IC of the ATE, $\varphi(\bm{z}) - \tau_p$, implies an efficient estimator $\hat{\tau}_p^* \equiv n^{-1} \sum_{i=1}^n \hat{\varphi}(\bm{z}_i)$, known as the augmented inverse propensity weighted (AIPW) estimator \citep{Robins1994}. We recommend the AIPW estimator in the current context since $\hat{\Theta}_p$ and $\hat{\Psi}_s$ are locally insensitive to small perturbations in $\hat{\tau}_p$ about $\hat{\tau}_p^*$. To see why, consider the partial derivative
\begin{align*}
    \frac{\partial \hat{\Theta}_{p}}{\partial \hat{\tau}_p} = \frac{\partial}{\partial \hat{\tau}_p} n^{-1}\sum_{i=1}^n \left[\{\hat{\varphi}(\bm{z}_i)-\hat{\tau}_p\}^2 - \{\hat{\varphi}(\bm{z}_i)-\hat{\tau}(\bm{x}_i)\}^2\right] = -2(\hat{\tau}_p^* - \hat{\tau}_p),
\end{align*}
which is zero at $\hat{\tau}_p = \hat{\tau}_p^*$. This orthogonality means that uncertainty in the AIPW estimator can be ignored when estimating $\Theta_p$, see \cite{Vermeulen2015} for details.

Like $\phi_s(\bm{z})$, $\Phi_s(\bm{z})$ degenerates to $\Phi_s(\bm{z})=0$ when $\Psi_{s}=0$ or when $\Psi_{s}=1$, i.e. $\Theta_s=0$ or $\Theta_s=\Theta_p$. For this reason, the asymptotic normality of $\hat{\Psi}_s$, described in Theorem \ref{asym_theorem2}, holds only for $\Psi_s \in (0,1)$, i.e. when covariates in $s$ account for some, but not all, heterogeneity. Similar estimators for $\log(\Theta_s)$ and $\logit(\Psi_s)$ are derived in Appendix \ref{TE-VIMscales}, which may be used to construct alternative bounded estimators $\hat{\Theta}^*_s > 0$ and $\hat{\Psi}^*_s \in (0,1)$.

\begin{theorem}
\label{asym_theorem2}
Assume that the conditions in Theorem \ref{asym_theorem} are satisfied, $\Theta_p > 0$, and there exists $\delta \in (0,\infty)$ such that $||\hat{\tau}(\bm{X}) - \hat{\tau}_p^*|| < \delta$. Then $\hat{\Psi}_{s}$, with the ATE estimated by $\hat{\tau}_p = \hat{\tau}_p^*$, is asymptotically linear with IC, $\Phi_s(\bm{Z})$, and hence $\hat{\Psi}_{s}$ converges to $\Psi_{s}$ in probability, and for $\Psi_{s} \in (0,1)$ then $n^{1/2}(\hat{\Psi}_{s}-\Psi_{s})$ converges in distribution to a mean-zero normal random variable with variance $||\Phi_s(\bm{Z})||^2$.
\end{theorem}

\subsubsection{Plug-in estimation}

A common framework for constructing debiased estimators is through targeted maximum likelihood estimators (TMLEs) \citep{VanDerLaan2016}. TMLEs are `plug-in', in that they are defined through estimand mappings, e.g. $\Theta_s:\mathcal{M}\mapsto [0,\infty)$ evaluated at a distribution estimate $\hat{P}_n\in\mathcal{M}$. Despite the apparent similarity of $\hat{\Theta}_s$ with the representation $\Theta_s = \E\left[\{\varphi(\bm{Z}) - \tau_s(\bm{X})\}^2 - \{\varphi(\bm{Z}) - \tau(\bm{X})\}^2\right]$, the estimators $\hat{\Theta}_{s}, \hat{\Theta}_{p}, \hat{\Psi}_{s}$ are not plug-in. This is evident from the fact that $\hat{\Theta}_{s}$ may take negative values, but such codomain violations are not possible for plug-in estimators. We emphasize this point since \cite{Williamson2021} use `plug-in' to refer to representation similarities, and their $\Omega_s$ estimator is also not plug-in in the estimand mapping sense. Moreover, TMLEs for $\Theta_s$ are challenging, since the targeting step must target compatible estimators for $\mu(\bm{x}_i)$ and $\mu_s(\bm{x}_i)$ simultaneously. The VTE does not suffer this issue, with TMLEs proposed by \cite{Levy2021}. TMLEs for TE-VIMs are investigated by \cite{Li2023}.

\subsubsection{Algorithms}

The estimators $\hat{\Theta}_{s}, \hat{\Theta}_{p}$, and $\hat{\Psi}_{s}$ are indexed by the choice of pseudo-outcome and CATE estimators. Generally, we are unrestricted in the choice of CATE metalearner, and outcome and propensity score learners. We propose two Algorithms based on the T- and DR-learners, with and without sample splitting. In Algorithm noSS substeps marked (A) and (B) refer to the T- and DR-learners respectively. Where the algorithms require models to be `fitted', any suitable regression method/learner can be used.

Both algorithms return pseudo-outcome and CATE estimates, $\{\hat{\varphi}_i\}_{i=1}^n,\{\hat{\tau}_i\}_{i=1}^n,$, and $\{\hat{\tau}_{s,i}\}_{i=1}^n$, which can be used to obtain $\hat{\tau}_p = n^{-1} \sum_{i=1}^n \hat{\varphi}_i $ and the uncentered ICs $\hat{\phi}_{i, s} = \{\hat{\varphi}_i-\hat{\tau}_{s,i}\}^2 - \{\hat{\varphi}_i-\hat{\tau}_{i}\}^2$ and $\hat{\phi}_{i, p} = \{\hat{\varphi}_i-\hat{\tau}_p\}^2 - \{\hat{\varphi}_i-\hat{\tau}_{i}\}^2$ which imply the estimators $\hat{\Theta}_s = n^{-1} \sum_{i=1}^n \hat{\phi}_{i, s}$, $\hat{\Theta}_p = n^{-1} \sum_{i=1}^n \hat{\phi}_{i, p}$, and $\hat{\Psi}_s = \hat{\Theta}_s/\hat{\Theta}_p$, with variances respectively estimated by $n^{-2} \sum_{i=1}^n \left(\hat{\phi}_{i, s} - \hat{\Theta}_s\right)^2$, $n^{-2} \sum_{i=1}^n \left(\hat{\phi}_{i, p} - \hat{\Theta}_p\right)^2$, and $(n\hat{\Theta}_p)^{-2} \sum_{i=1}^n \left(\hat{\phi}_{i, s} - \hat{\Psi}_s\hat{\phi}_{i, p}\right)^2$. The algorithms differ in their CATE metalearner and use of sample splitting. Comparing Algorithms SS-A and SS-B, the DR-learner requires additional cross-fitting because it is trained on pseudo-outcomes estimates that are learned from a separate sample. As such, Algorithm SS-B has $\mathcal{O}(K^2)$ regression operations compared with $\mathcal{O}(K)$ for SS-A.

\textbf{Algorithm noSS: No sample splitting}
\begin{enumerate}[label=(\arabic*)]
\item Fit $\hat{\mu}(.,.)$ and $\hat{\pi}(.)$. Use these fitted models to obtain $\hat{\varphi}_i \equiv\hat{\varphi}(z_i)$.
\item (A) Use the model for $\hat{\mu}(.,.)$ from Step 1, to obtain $\hat{\tau}(\bm{x}) \equiv\hat{\mu}(1,\bm{x})-\hat{\mu}(0,\bm{x})$. Or (B)  Fit $\hat{\tau}(.)$ by regressing $\hat{\varphi}(\bm{Z})$ on $\bm{X}$. After doing (A) or (B), use the fitted models to obtain $\hat{\tau}_i \equiv \hat{\tau}(\bm{x}_i)$.
\item Fit $\hat{\tau}_s(.)$ by regressing $\hat{\tau}(\bm{X})$ on $\bm{X}_{-s}$. Use the fitted model to obtain $\hat{\tau}_{s,i} \equiv\hat{\tau}_s(\bm{x}_i)$.
\item Optionally repeat Step 3 for other covariate sets of interest.
\end{enumerate}

\textbf{Algorithm SS-A: Sample splitting with the T-Learner}
\begin{enumerate}[label=(\arabic*)]
\item Split the data into $K\geq 2$ folds. \textbf{For} each fold $k$: Fit $\hat{\mu}(.,.)$ and $\hat{\pi}(.)$ using the data set excluding fold $k$. Use these fitted models to obtain $\hat{\varphi}_i \equiv\hat{\varphi}(\bm{z}_i)$ and $\hat{\tau}_{i} \equiv\hat{\mu}(1,\bm{x}_i)-\hat{\mu}(0,\bm{x}_i)$ for $i$ in fold $k$.
\item Fit $\hat{\tau}_s(.)$ by regressing $\hat{\mu}(1,\bm{X})-\hat{\mu}(0,\bm{X})$ on $\bm{X}_{-s}$ using the data excluding fold $k$. Use the fitted model to obtain $\hat{\tau}_{s,i} \equiv\hat{\tau}_s(\bm{x}_i)$ for $i$ in fold $k$.
\item Optionally repeat Step 2 for other covariate sets of interest. \textbf{End for.}
\end{enumerate}

\textbf{Algorithm SS-B: Sample splitting with the DR-Learner}
\begin{enumerate}[label=(\arabic*)]
% \item Split the data into $K\geq 3$ folds.
\item  Split the data into $K\geq 3$ folds. \textbf{For} each pair of folds $j\neq k$: Fit $\hat{\mu}(.,.)$ and $\hat{\pi}(.)$ using the data set that excludes both folds $j$ and $k$. Use these fitted models to obtain $\hat{\varphi}^{(k)}_i \equiv\hat{\varphi}(\bm{z}_i)$ for $i$ in fold $j$, and $\hat{\varphi}^{(j)}_i \equiv\hat{\varphi}(\bm{z}_i)$ for $i$ in fold $k$. \textbf{End for.}
\item \textbf{For} each fold $k$: Fit $\hat{\tau}(.)$ by regressing $\hat{\varphi}^{(k)}(\bm{Z})$ on $\bm{X}$ using the data excluding fold $k$. Use the fitted models to obtain $\hat{\tau}_i \equiv \hat{\tau}(\bm{x}_i)$ for $i$ in fold $k$.
\item Obtain $\hat{\varphi}_i \equiv (K-1)^{-1}\sum_{j\neq k} \hat{\varphi}^{(j)}(\bm{z}_i)$ for $i$ in fold $k$.
\item Fit $\hat{\tau}_s(.)$ by regressing $\hat{\tau}(\bm{X})$ on $\bm{X}_{-s}$ using the data excluding fold $k$. Use the fitted model to obtain $\hat{\tau}_{s,i} \equiv\hat{\tau}_s(\bm{x}_i)$ for $i$ in fold $k$.
\item Optionally repeat Step 4 for other covariate sets of interest. \textbf{End for.}
\end{enumerate}

\section{Simulation study}
\label{sect:simulation}

We compared all Algorithms ($K=8$ folds) under three data generating processes (DGPs).
For each, generalized additive models (GAMs) were used to estimate $\mu(a,\bm{x})$, $\pi(\bm{x})$, $\tau_s(\bm{x})$, and in the case of the DR-learner, $\tau(\bm{x})$. GAMs are flexible spline smoothing models implemented in the \verb|mgcv| R package \citep{Wood2016}, and for DGPs 1 and 2, $(X_1, X_2)$ interaction terms were included. Propensity score models were additive on the logit scale.

\textbf{DGP 1}: We generated $500$ datasets for each size $n\in \{500, 1000, 2000, 3000, 4000, 5000\}$ according to $X_1,X_2 \sim \text{Uniform}(-1,1)$, $A \sim \text{Bernoulli}\{\text{expit}(- 0.4X_1 + 0.1X_1 X_2)\}$, $\tau^{(1)}(\bm{X}) = X_1^3 + 1.4 X_1^2 + 25 X_2^2/9$, and $Y \sim \mathcal{N}_1(X_1X_2 + 2X_2^2 - X_1 + A\tau^{(1)}(\bm{X}), 1)$ where $\mathcal{N}_p(0, \Sigma)$ denotes a $p$-dimensional normal variable with mean $\mu$ and covariance matrix $\Sigma$. In this case $\tau_p = 1.39$ and $\Theta_p= 1.003$. Since we consider only two covariates, the LOO and KOI TE-VIM modes are equivalent, with $\Theta_1 = 0.32$, $\Theta_2 = 0.69$, $\Psi_1 = 0.32$, and $\Psi_2 = 0.68$.

\textbf{DGP 2}: The setup in DGP 1, but with $\tau^{(1)}(\bm{X})$ replaced with $\tau^{(2)}(\bm{X}) = \tau^{(1)}(\bm{X}) / 10$. In this way, the relative importance of $X_1, X_2$ are unchanged, but the overall effect size and heterogeneity is much smaller. This results in $\tau_p = 0.139$, $\Theta_p= 0.01003$, $\Theta_1 = 0.0032$ and $\Theta_2 = 0.0069$, but the scaled TE-VIMs $\Psi_1 = 0.32$ and $\Psi_2 = 0.68$ are the same as DGP 1.

\textbf{DGP 3}: We generated $500$ datasets with $n=5000$ according to
\begin{align*}
(X_1, X_2), (X_3, X_4), (X_5, X_6) &\sim \mathcal{N}_2\left(0, \begin{pmatrix}
1 & 0.5 \\
0.5 & 1 
\end{pmatrix}
\right),
\end{align*}
$A \sim \text{Bernoulli}\{\text{expit}(- 0.4X_1 + 0.1X_1 X_2  + 0.2 X_5)\} $, $\tau^{(3)}(\bm{X}) = X_1 + 2 X_2 + X_3$, $Y \sim \mathcal{N}_1(X_3 - X_6 + A\tau^{(3)}(\bm{X}), 3)$. In this case $\tau_p = 0$ and $\Theta_p= 8$ but the LOO and KOI TE-VIM modes are not equivalent (see Table \ref{dgp_2}). Under the KOI mode, some importance is assigned to $X_4$ due to its correlation with $X_3$, also greater importance is assigned to $X_1$ versus $X_3$ due to the correlation of $X_1$ with $X_2$. The LOO mode assigns little importance to $X_1, X_3$ since they are correlated with $X_2, X_4$ respectively. Shapley values represent a compromise between these modes, but require $2^p = 64$ TE-VIMs to be evaluated. For this reason we compared only the LOO and KOI modes in our simulation.

\begin{table}[htb]
\caption{True TE-VIM values for DGP 3. For the KOI mode we report $\Theta_p-\Theta_s$. Scaled TE-VIMs are obtained by dividing each value by $\Theta_p = 8$. The Shapley values sum to $\Theta_p$ by design. }
\label{dgp_2}
\centering
\begin{tabular}{|l|l|l|l|}
\hline
Target covariate & Leave-one-out & Keep-one-in & Shapley\\ \hline
$X_1$ & $0.75$ & $4$ & $3.375$\\ \hline
$X_2$ & $3$ & $6.25$ & $4.625$ \\ \hline
$X_3$ & $0.75$ & $1$ & $0.875$ \\ \hline
$X_4$ & $0$ & $0.25$ & $0.125$ \\ \hline
$X_5$ & $0$ & $0$ & $0$ \\ \hline
$X_6$ & $0$ & $0$ & $0$ \\ \hline
\end{tabular}
\end{table}

\subsection{Results}

\begin{figure}[htbp]
\centering
\includegraphics[width=\linewidth]{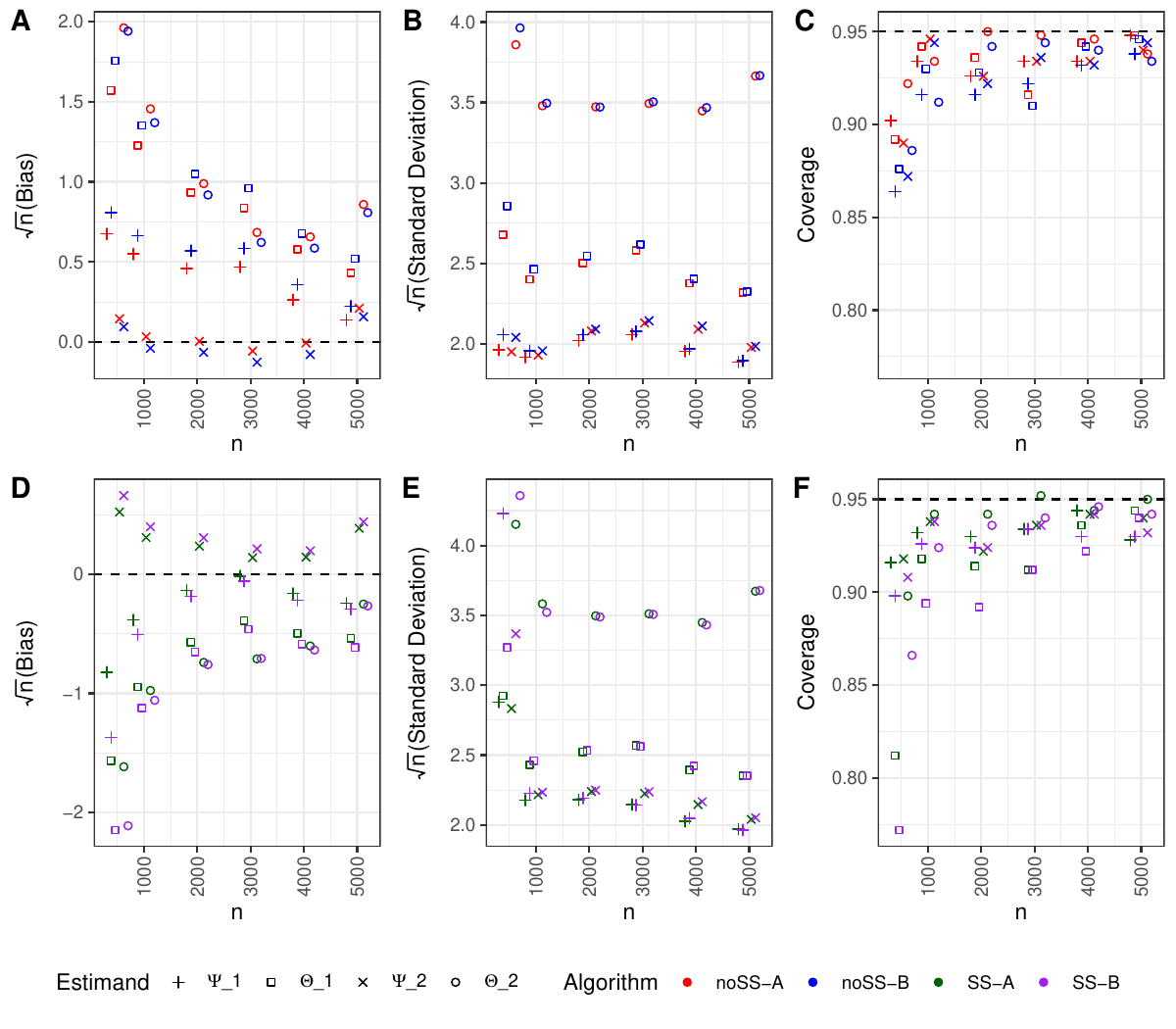}
\caption{Bias (A, D), empirical standard deviation (B, E) and coverage (C, F) for estimators from DGP 1. Dashed lines indicate zero bias, and nominal 95\% CI coverage. For readability, a small amount of `jitter' has been added to the sample size, $n$. In order to present scaled and unscaled TE-VIMs together, the standard deviation and bias of scaled TE-VIMs has been multiplied by the true VTE. Note that the bias and variance scales differ between sub-plots A,B and D,E.}
\label{sim_results_dgp1}
\end{figure}

For each dataset, $\Theta_s$ and $\Psi_s$ were estimated, along with their standard errors and Wald based (95\%) confidence intervals (CIs). For DGPs 1 and 2, we also report the empirical probability that TE-VIMs correctly rank $X_2$ as more important than $X_1$. Figure \ref{sim_results_dgp1} shows bias, variance, and coverage plots for DGP 1. Additional plots for DGP 2 and DGP 3 are in Appendix \ref{appendix:additional_plots}.

For all DGPs, we see that TE-VIM estimators which do not use sample splitting tend to overestimate TE-VIMs (positive bias), whilst sample splitting estimators tend to underestimate TE-VIMs (negative bias). For DGP 1, we observe that, in small samples, DR-learner based algorithms (-B) produce larger bias, variance, and reduced CI coverage, than their T-learner counterparts (-A). Moreover, the scaled TE-VIM estimators tend to have smaller bias and variance than TE-VIM estimators. This trend appears reversed in the low-heterogeneity regime (DGP 2) when sample splitting is used. We believe this is due to extreme inverse weighting in the VTE estimate $\hat{\Theta}_p$, which appears in the denominator of $\hat{\Psi}_1$ and $\hat{\Psi}_2$.

For DGP 1, all algorithms recover the correct ranking with a high degree of accuracy. For DGP 2, this accuracy is reduced and conclusions based on scaled and unscaled TE-VIMs do not always agree, with the latter generally being more correct when cross-fitting is used (see Appendix \ref{appendix:additional_plots}). For a given dataset, the ranking of scaled and unscaled TE-VIMs differs only when $\hat{\Theta}_p < 0$. Therefore, we recommend that scaled TE-VIMs are only used when sensible VTE estimates are obtained, though TE-VIMs are also scientifically less relevant when there is little heterogeneity to account for.

In DGP 3 we observe that null importance does not seem to affect estimator bias, but does lead to reduced estimator standard deviations, as expected from theory, and decreased CI coverage. This phenomenon is especially clear when examining covariate $X_4$, which has, in truth, null importance under the LOO mode, but not under the KOI mode. For the $X_4$ LOO TE-VIM estimators we observe low variance and low CI coverage, whereas for the $X_4$ KOI TE-VIM estimators we see higher variance and closer to nominal coverage. 

\section{Applied example: AIDS clinical trial}
\label{sect:example}

The AIDS Clinical Trials Group Protocol 175 (ACTG175) \citep{Hammer1996} considers 2139 HIV patients with CD4 T-cell count between 200 and 500$mm^{-3}$ randomized to 4 treatment groups: (i) zidovudine (ZDV); (ii) didanosine (ddI); (iii) ZDV+ddI; (iv) ZDV+zalcitabine. We compare groups (ii) and (iii) $(A=0,1)$, with 561 and 522 patients respectively. We consider CD4 count at 20±5 weeks as a continuous outcome $Y$ and 12 baseline covariates, 5 continuous: age, weight, Karnofsky score, CD4/CD8 count; and 7 binary: sex, homosexual activity (y/n), race (white/non-white), symptomatic (y/n), intravenous drug use history(y/n), hemophilia (y/n), and antiretroviral history (experienced/naive).

TE-VIMs for each covariates were estimated using all algorithms with $K=10$ folds (around 10 folds is typical for cross-fitting procedures). A constant propensity score of $522 / 1083 \approx 0.48$ was used, since treatment is randomized. Fitted models for the outcome and CATEs were obtained using the `discrete' Super Learner \citep{VanDerLaan2007}, an ensemble learning method, which selects the regression algorithm in a `learner library' that minimizes some cross-validated risk. We used the \verb|SuperLearner| R package implementation of this algorithm with 10 cross-validation folds, mean-squared-error loss, and a learner library containing various routines (\verb|glm|, \verb|glmnet|, \verb|gam|, \verb|xgboost|, \verb|ranger|). Similar results are obtained when Super Learner regression is ablated and replaced with GAMs (see Appendix \ref{supp:gam}).

\begin{figure}[htbp]
\centering
\includegraphics[width=\linewidth]{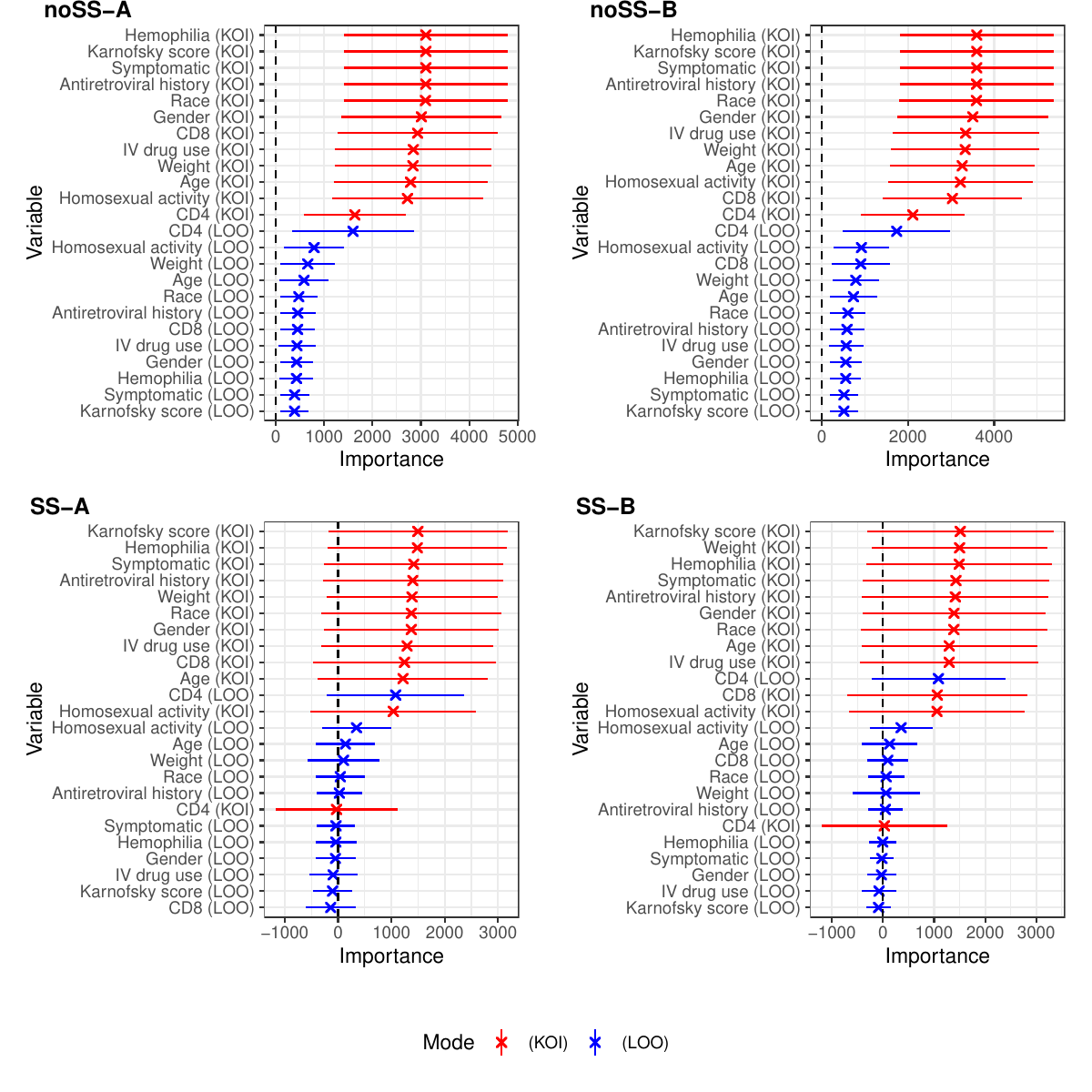}
\caption{TE-VIM estimates $\hat{\Theta}_s$ from the ACTG175 study using each of the proposed Algorithms. Error bars indicate 95\% CIs. In each plot, covariates are sorted according to their TE-VIM point estimate. Dashed lines indicate no importance. For the KOI mode, the TE-VIM represents the importance of the complement variable set, i.e. low-values denote high-importance of the KOI covariate.} 
\label{results_plot_unscaled}
\end{figure}

\begin{figure}[htbp]
\centering
\includegraphics[width=\linewidth]{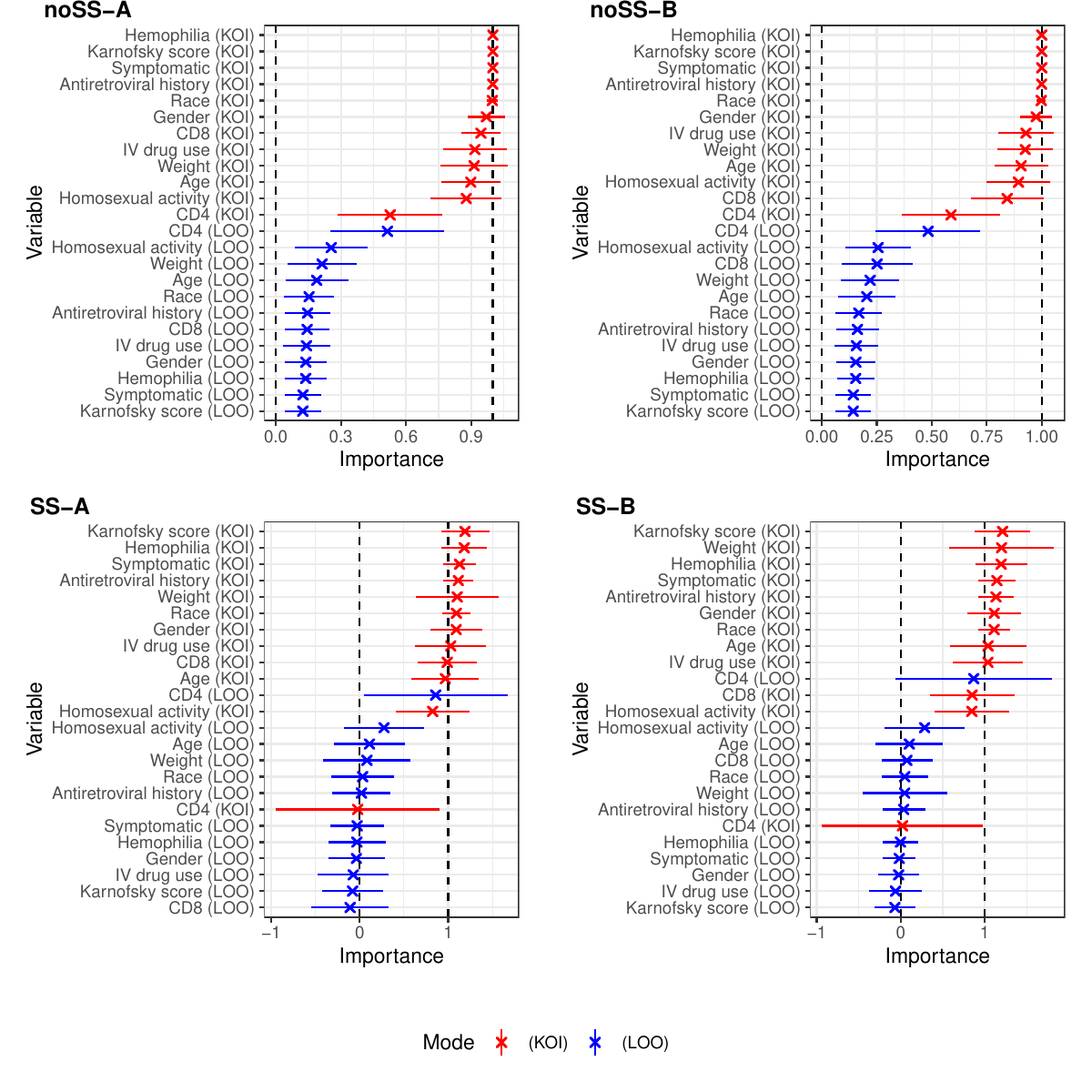}
\caption{Scaled TE-VIM estimates $\hat{\Psi}_s$ from the ACTG175 study using each of the proposed Algorithms. Error bars indicate 95\% CIs. In each plot, covariates are sorted according to their TE-VIM point estimate. Dashed lines indicate the $[0,1]$ support of the scaled TE-VIM. For the KOI mode, the TE-VIM represents the importance of the complement variable set, i.e. low-values denote high-importance of the KOI covariate.} 
\label{results_plot_scaled}
\end{figure}

AIPW estimates of the ATE using pseudo-outcomes from Algorithms noSS, SS-A, and SS-B were similar, respectively: 28.2$mm^{-3}$ (CI: 14.0, 42.3; p$<$0.01); 28.4$mm^{-3}$ (CI: 13.8, 42.9; p$<$0.01); 27.9$mm^{-3}$ (CI: 13.3, 42.5; p$<$0.01), where all CIs are reported at 95\% significance and  p-values are of Wald type. VTE estimates differed substantially between algorithms with/without cross-fitting. With Algorithms noSS-A and -B returning estimates: 3100$mm^{-6}$ (CI: 1410, 4790; p$<$0.01) and 3600$mm^{-6}$ (CI: 1810, 5380; p$<$0.01), and for Algorithms SS-A and -B: 1260$mm^{-6}$ (CI: -425, 2940; p$=$0.14) and 1250$mm^{-6}$ (CI: -580, 3080; p$=$0.18). It is helpful to also consider the square root of the VTE estimates, which is on the same scale as the ATE. These are 55.7, 60.0, 35.5, and 35.3$mm^{-3}$ for Algorithms noSS-A, -B, SS-A, and -B respectively. Based on the VTE CIs from Algorithms SS-A and -B, low treatment effect heterogeneity is a concern in this analysis. Figures \ref{results_plot_unscaled} and \ref{results_plot_scaled} show unscaled and scaled TE-VIM estimates using LOO and KOI modes. All Algorithms rank CD4 count and homosexual activity as the most important covariates, with CD8 count also ranked highly by Algorithm SS-B under the KOI mode. We also observe (i) that standard errors are small for unimportant covariates, as expected due to the importance testing issues in Section \ref{sect:import}; (ii) Algorithms SS-A and -B produce scaled TE-VIMs point estimates outside of $(0, 1)$ more often than noSS-A and -B; and (iii) Algorithms noSS-A and -B do not produce scaled or unscaled TE-VIM point estimates with 95\% CIs overlapping zero.

\section{Related work and extensions}
\label{sect:extend}

Here we discuss VIMs for the OTR and TE-VIMs for continuous treatments. Discussion on treatment effect scales, linear CATE projections \citep{Boileau2022} and treatment effect cumulative distribution functions \citep{Levy2018} are in Appendix \ref{appendix:additional_discussion}.

\subsection{Optimal treatment rule - VIMs}
\label{sect:otr_vims}

Although TE-VIMs capture the importance of variables in explaining the CATE, this should not be confused with the importance of those variables in determining the OTR $d(\bm{x})\equiv\mathbb{I}\{\tau(\bm{x})>0\}$, where $\mathbb{I}(.)$ is an indicator and, w.l.o.g., a greater outcome is preferred. For instance, treatment could be uniformly beneficial, in which case the OTR is to always treat, despite possible treatment effect heterogeneity. In such settings, analysis using TE-VIMs may provide useful insight to further improve therapies. Faced with OTR heterogeneity, one might consider nonparametric VIMs related to the OTR, e.g. by considering the risk $f \mapsto -\E\{Y^{f(X)}\}$, where $f: \mathbb{R}^p \mapsto \{0, 1\}$ is a policy. This risk implies the OTR-VIM $\Gamma^*_s  \equiv \E\{Y^{d(\bm{X})} - Y^{d^*_s(\bm{X})}\} \geq 0$ where $d^*_s(\bm{x}) \equiv\mathbb{I}\{\tau_s(\bm{x})>0\}$ is the OTR given $\bm{X}_{-s}$ \citep{Zhang2015, Williamson2021_general}. Thus, OTR-VIMs compare the OTR with suboptimal policies that are optimal within a restricted set of policies.

Under standard causal assumptions (consistency, positivity, exchangeability), the OTR-VIM is identified by $\Gamma^*_s = \E\left[\tau(\bm{X})\{d(\bm{X})-d^*_s(\bm{X})\}\right]$. Unlike TE-VIMs, $\Gamma^*_s$ is not pathwise differentiable without additional assumptions e.g. that the OTR is insensitive to small changes in $P_0$ \citep{Luedtke2016}, with the pathwise derivative usually used to construct efficient estimators. Analogous to $\Theta_s$ and $\Omega_s$, one could alternatively define OTR-VIMs using the risk $f\mapsto \E[\{d(\bm{X}) - f(\bm{X})\}^2]$, which implies the VIM $\Gamma_s \equiv \E[d_s(\bm{X})\{1-d_s(\bm{X})\}]$ where $d_s(\bm{x}) \equiv \E\{d(\bm{X})|\bm{X}_{-s}=\bm{x}_{-s}\}=Pr\{\tau(\bm{X}) >0 |\bm{X}_{-s}=\bm{x}_{-s}\}$. Note that $d(\bm{X})\in\{0,1\}$ and $d_s(\bm{X})\in[0,1]$. Like $\Theta_s$ and $\Omega_s$, $\Gamma_s \in [0,0.25]$ is invariant to linear transformations of the outcome, however, like $\Gamma^*_s$, additional assumptions on the OTR are required for pathwise differentiability of $\Gamma_s$.

\subsection{Continuous treatments}
\label{ext:continuous}

Continuous analogues of the CATE based on linear model projections are proposed by \cite{Hines2023}. In particular, $\lambda(\bm{x}) \equiv \Cov(A,Y|\bm{X}=\bm{x}) / \Var(A|\bm{X}=\bm{x})$ is well defined when $A$ is continuous, and identifies the CATE under standard causal assumptions (consistency, positivity, exchangeability) when $A$ is binary. Appealing to the risk $f\mapsto ||\lambda(\bm{X}) - f(\bm{X})||^2$, one might extend the ATE, VTE, and TE-VIMs to continuous exposures using the estimands: $\E\{\lambda(\bm{X})\}$, $\Var\{\lambda(\bm{X})\}$, and $\E[\Var\{\lambda(\bm{X})|\bm{X}_{-s}\}] / \Var\{\lambda(\bm{X})\}$, which identify their CATE counterparts when $A$ is binary. ICs for these estimands are obtained by replacing the pseudo-outcome $\varphi(\bm{z})$ in the current work, with
\begin{align*}
\left[y-\mu(\bm{x})-\lambda(\bm{x})\{a-\pi(\bm{x})\} \right] \frac{a-\pi(\bm{x})}{\Var(A|\bm{X}=\bm{x})} + \lambda(\bm{x})
\end{align*}
which reduces to $\varphi(\bm{z})$ when $A$ is binary. See Appendix \ref{supp:continuo} for details. 

\section{Conclusion}

We propose TE-VIMs, which capture the importance of variable subsets in explaining the CATE, and complement the VTE as a global heterogeneity measure \citep{Levy2021}. We derive efficient TE-VIM estimators that are amenable to machine learning of working models, thus, unlike existing proposals using causal forests, are not tied to specific algorithms \citep{Athey2019}.
In studies where heterogeneous treatment effects are of primary interest, we recommend using TE-VIMs to select stratification variables for subgroup analyses, to support variable selection decisions for CATE/ OTR models, or to rank effect modifiers. In studies where the main goal is to estimate the ATE, we recommend that VTE inference should also form part of the analysis, since the ATE and VTE can be used to inform on the marginal probability of adverse CATEs (see Appendix \ref{supp:tecdf}). TE-VIMs may then form part of a secondary analysis when large treatment effect heterogeneity cannot be ruled out. We recommend that domain specific knowledge is used to select potential effect modifiers for VTE and TE-VIM analyses. We have outlined several frameworks for choosing the sets of variables to be included for the comparisons needed in the TE-VIM algorithm, such as leave-one-out or keep-one-in. Finally, we caution users against over-interpreting TE-VIM p-values due to the difficulty of testing the zero importance null and due to the multiplicity problem when the number of covariate subsets is large. The latter problem may be somewhat attenuated using a Bonferroni approach. Any suggested covariate that may explain heterogeneity of treatment effects should be confirmed using independent data.

\section*{Supplementary Materials}

Replication code is available at \url{https://github.com/ohines/tevims}. AIDS Clinical Trials Group Protocol 175 (ACTG175) data \cite{Hammer1996} is available on CRAN at \url{https://CRAN.R-project.org/package=speff2trial}.

\bibliography{refs}
\bibliographystyle{apalike}

\pagebreak

\appendix

\input{supplement.tex}

\end{document}

%% file: supplement.tex
\section{Derivation of Efficient Influence Curve}
\label{ic_deriv_append}

We adopt the IC derivation formalism given in \cite{Hines2021}. Specifically we let $P_0$ denote the true distribution of $(Y,A,\bm{X})$ and let $\tilde{P}$ denote a point mass at $(\tilde{y},\tilde{a},\tilde{\bm{x}})$. We further denote the parametric submodel $P_t = t\tilde{P} + (1-t)P_0$ where $t\in[0,1]$ is a scalar parameter, and we let $\partial_t$ denote an operator such that for some function of $f(t)$, $\partial_t f(t)\equiv \frac{d f(t)}{dt}|_{t=0}$.

Our goal is to derive an IC for $L^*_{P_0}\{h\} \equiv \E_{P_0}[\{\tau(\bm{X}) - h(\bm{X})\}^2]$, for a known function $h:\mathbb{R}^p\mapsto\mathbb{R}$, which we will connect to ICs for $\Theta_s$ and $\Psi_s$.

We make use of the following lemma, which we demonstrate later in the proof. Letting $g_P(X)$ denote some functional of $P$, then
\begin{align}
&\partial_t \E_{P_t}\{g_{P_t}(\bm{X})|\bm{X}_{-s}=x_{-s}\}  \nonumber \\ 
&= \frac{\tilde{f}(\bm{x}_{-s})}{f(\bm{x}_{-s})} \left[ g_{P_0}(\tilde{\bm{x}}) - \E_{P}\{g_{P_0}(\bm{X})|\bm{X}_{-s}=\bm{x}_{-s}\}\right] + \E_{P_0}\{\partial_t g_{P_t}(\bm{X})|\bm{X}_{-s}=\bm{x}_{-s}\} \label{quick_lemma}
\end{align}
where $\tilde{f}(.)$ and $f(.)$ denote the marginal `densities' of $\bm{X}_{-s}$ under $\tilde{P}$ and $P_0$ respectively, which are w.l.o.g. absolutely continuous w.r.t. to a dominating measure. In practice this expression means that for discrete $\bm{X}_{-s}$ then $f(.)$ is a probability mass function and $\tilde{f}(.)$ is an indicator function. Similarly for continuous $\bm{X}_{-s}$ then $f(.)$ is a probability density function and $\tilde{f}(.)$ is a Dirac delta function. In both cases $\tilde{f}(\bm{x}_{-s})$ is a probability point mass, which is zero when $\tilde{\bm{x}}_{-s} \neq \bm{x}_{-s}$.

It follows immediately from \eqref{quick_lemma} that,
\begin{align}
\partial_t \E_{P_t}\{g_{P_t}(\bm{X})\} &= g_{P_0}(\tilde{\bm{x}}) - \E_{P_0}\{g_{P_0}(\bm{X})\} + \E_{P_0}\{\partial_t g_{P_t}(\bm{X})\} \label{quick_lemma2}
\end{align}
For the function $g_{P_t}(\bm{x}) = \{\tau_t(\bm{x}) - h(\bm{x})\}^2$, where $\tau_t(\bm{x})$ represents $\tau(\bm{x})$ under $P_t$, we obtain
\begin{align}
\partial_t L^*_{P_t}\{h\} = \{\tau(\tilde{\bm{x}}) - h(\tilde{\bm{x}})\}^2 - L^*_{P_0}\{h\} + 2 \E_{P_0}[\{\tau(\bm{x}) - h(\bm{x})\}\partial_t \tau_t(\bm{x})]
\end{align}
we use \eqref{quick_lemma} and the fact that $\tau(\bm{x})=\mu(1,\bm{x})-\mu(0,\bm{x})$ to show that,
\begin{align*}
\partial_t \tau_t(\bm{x}) &= \frac{\tilde{f}(\bm{x})}{f(\bm{x})}\{\tilde{y}-\mu(\tilde{a},\bm{x})\}\frac{\tilde{a} - \pi(\bm{x})}{\pi(\bm{x})\{1-\pi(\bm{x})\}}
\end{align*}
hence we obtain the IC
\begin{align}
\partial_t L^*_{P_t}\{h\} &= \{\tau(\tilde{\bm{x}}) - h(\tilde{\bm{x}})\}^2 - L^*_{P_0}\{h\} + 2 \{\tau(\tilde{\bm{x}}) - h(\tilde{\bm{x}})\}\{\tilde{y}-\mu(\tilde{a},\tilde{\bm{x}})\}\frac{\tilde{a} - \pi(\tilde{\bm{x}})}{\pi(\tilde{\bm{x}})\{1-\pi(\tilde{\bm{x}})\}} \\
&= \{\tau(\tilde{\bm{x}}) - h(\tilde{\bm{x}})\}^2 - L^*_{P_0}\{h\} + 2 \{\tau(\tilde{\bm{x}}) - h(\tilde{\bm{x}})\}\{\varphi(\tilde{\bm{z}}) - \tau(\tilde{\bm{x}})\}
\end{align}
Completing the square of the expression above gives
\begin{align}
\partial_t L^*_{P_t}\{h\} &= \{\varphi(\tilde{\bm{z}}) - h(\tilde{\bm{x}})\}^2 - \{\varphi(\tilde{\bm{z}}) - \tau(\tilde{\bm{x}})\}^2 - L^*_{P_0}\{h\}
\end{align}
When replicating this proof, it is useful to note that for an arbitrary function $w(\bm{x})$
\begin{align*}
\E_{P_0}\left\{ \frac{\tilde{f}(\bm{X})}{f(\bm{X})} w(\bm{X}) \right\} &= w(\tilde{\bm{x}})
\end{align*}

\subsection{Proof of Lemma in \eqref{quick_lemma}}
To demonstrate \eqref{quick_lemma} we write the lefthand side as
\begin{align*}
\partial_t \int g_{P_t}(\bm{x}^*) dP_{t,\bm{X}_s|\bm{x}_{-s}}(\bm{x}^*_s) &= \int g_{P}(\bm{x}^*) \partial_t dP_{t,\bm{X}_s|\bm{x}_{-s}}(\bm{x}^*_s) + \int \{\partial_t g_{P_t}(\bm{x}^*)\} dP_{\bm{X}_s|\bm{x}_{-s}}(\bm{x}^*_s)
\end{align*}
where $dP_{t,\bm{X}_s|\bm{x}_{-s}}(.)$ is the conditional distribution of $\bm{X}_s$ given $\bm{X}_{-s}=\bm{x}_{-s}$ under the parametric submodel and $\bm{x}^*_{-s} = \bm{x}_{-s}$. The second integral on the righthand side recovers the final term in \eqref{quick_lemma}. Hence the lemma follows once we show that
\begin{align*}
 \partial_t dP_{t,\bm{X}_s|\bm{x}_{-s}}(\bm{x}^*_s) = \frac{\tilde{f}(\bm{x}_{-s})}{f(\bm{x}_{-s})} \{ d\tilde{P}_{\bm{X}_s}(\bm{x}^*_s) - dP_{\bm{X}_s|\bm{x}_{-s}}(\bm{x}^*_s)\}
\end{align*}
To do so, let $\mu$ denote a dominating measure and write
\begin{align*}
dP_{t,\bm{X}_s|\bm{x}_{-s}}(\bm{x}^*_s) &= f_{t,\bm{X}_s|\bm{x}_{-s}}(\bm{x}^*_s) d\mu(\bm{x}^*_s) \\
&= \frac{f_{t,\bm{X}}(\bm{x}^*)}{f_{t,\bm{X}_{-s}}(\bm{x}_{-s})} d\mu(\bm{x}^*_s)
\end{align*}
where $f_{t,\bm{X}}(.)$ and $f_{t,\bm{X}_{-s}}(.)$ denote the marginal densities of $\bm{X}$ and $\bm{X}_{-s}$ under the parametric submodel, $P_t$, i.e. they are the Radon-Nikodym derivatives w.r.t. $\mu$. Applying the quotient rule, we obtain
\begin{align*}
\partial_t dP_{t,\bm{X}_s|\bm{x}_{-s}}(\bm{x}^*_s) &= \frac{1}{f_{\bm{X}_{-s}}(\bm{x}_{-s})} \left[ \partial_t f_{t,\bm{X}}(\bm{x}^*) -
\frac{f_{\bm{X}}(\bm{x}^*)}{f_{\bm{X}_{-s}}(\bm{x}_{-s})}   \partial_t f_{t,\bm{X}_{-s}}(\bm{x}_{-s}) \right]d\mu(\bm{x}^*_s)
\end{align*}
We now evaluate the derivative parts. Since $\partial_t P_t = \tilde{P}-P$, the marginal density derivatives will have a similar structure, as shown in the first expression below, where $f_{\bm{X}}(.)$ and $\tilde{f}_{\bm{X}}(.)$ denote marginal densities of $\bm{X}$ under $\tilde{P}$ and $P$, with likewise for $\bm{X}_{-s}$
\begin{align*}
\partial_t dP_{t,\bm{X}_s|\bm{x}_{-s}}(\bm{x}^*) &=\frac{1}{f_{\bm{X}_{-s}}(\bm{x}_{-s})} \left[ \{\tilde{f}_{\bm{X}}(\bm{x}^*)-f_{\bm{X}}(\bm{x}^*)\} -
  \frac{f_{\bm{X}}(\bm{x}^*)}{f_{\bm{X}_{-s}}(\bm{x}_{-s})}   \{\tilde{f}_{\bm{X}_{-s}}(\bm{x}_{-s})-f_{\bm{X}_{-s}}(\bm{x}_{-s})\} \right]d\mu(\bm{x}^*_s)
\end{align*}
Since $\tilde{P}$ is a point mass, $\tilde{f}_{\bm{X}}(\bm{x}^*) = \tilde{f}_{\bm{X}_s}(\bm{x}^*_s)\tilde{f}_{\bm{X}_{-s}}(\bm{x}^*_{-s})$. Also $\bm{x}^*_{-s} = \bm{x}_{-s}$ hence,
\begin{align*}
\partial_t dP_{t,\bm{X}_s|\bm{x}_{-s}}(\bm{x}^*) &=\frac{\tilde{f}_{\bm{X}_{-s}}(\bm{x}_{-s})}{f_{\bm{X}_{-s}}(\bm{x}_{-s})} \left[ \tilde{f}_{\bm{X}_s}(\bm{x}^*_s)  -
  \frac{f_{\bm{X}}(\bm{x}^*) }{f_{\bm{X}_{-s}}(\bm{x}_{-s})}   \right]d\mu(\bm{x}^*_s) \\
  &= \frac{\tilde{f}_{\bm{X}_{-s}}(\bm{x}_{-s})}{f_{\bm{X}_{-s}}(\bm{x}_{-s})} \left[ \tilde{f}_{\bm{X}_s}(\bm{x}^*_s) -
    f_{\bm{X}_s|\bm{x}_{-s}}(\bm{x}^*_s)   \right]d\mu(\bm{x}^*_s)
\end{align*}
Thus, the result follows.

\section{Estimator Asymptotic Distributions}
\label{Appen_assymp}

In this Appendix we use a common empirical processes notation, where we define linear operators $P_0$ and $\mathbb{P}_n$ such that for some function $h(\bm{Z})$, $P_0\{ h(\bm{Z})\} \equiv \E \{h(\bm{Z})\}$ and $\mathbb{P}_n \{h(\bm{Z})\} \equiv n^{-1} \sum_{i=1}^n h(\bm{z}_i)$. To simplify notation we also largely omit function arguments, for example $\tau = \tau(\bm{X})$ with similar for $\hat{\tau},\tau_s,\hat{\tau}_s,\pi,\hat{\pi},\hat{\varphi}$. 

\subsection{Proof of Theorem \ref{asym_theorem}}

Define
\begin{align*}
\hat{\varphi}(\bm{z}) &\equiv\{y-\hat{\mu}(a,\bm{x})\}\frac{a-\hat{\pi}(\bm{x})}{\hat{\pi}(\bm{x})\{1-\hat{\pi}(\bm{x})\}} + \hat{\mu}(1,\bm{x}) - \hat{\mu}(0,\bm{x}) \\
\hat{\phi}_s(\bm{z}) &\equiv \{\hat{\varphi}(\bm{z})-\hat{\tau}_s(\bm{x})\}^2 - \{\hat{\varphi}(\bm{z})-\hat{\tau}(\bm{x})\}^2  - \hat{\Theta}_s^0
\end{align*}
where $\hat{\Theta}_s^0$ is an initial estimate of $\Theta_s$.  Without making any restrictions we write
\begin{align}
    \hat{\Theta}_s - \Theta_s &= (\mathbb{P}_n -P_0)\{\phi_{s}(\bm{Z})\} + R_n + H_n \\
    \hat{\Theta}_s &\equiv \hat{\Theta}_s^0 + \mathbb{P}_n\{\hat{\phi}_{s}(\bm{Z})\} \\
    R_n &\equiv \hat{\Theta}_s^0 - \Theta_s + P_0\{\hat{\phi}_{s}(\bm{Z})\} \\
    H_n &\equiv (\mathbb{P}_n -P_0)\{\hat{\phi}_{s}(\bm{Z}) - \phi_{s}(\bm{Z})\}.
\end{align}
We will show that the remainder term $R_n=o_P(n^{-1/2})$ and the empirical process term $H_n=o_P(n^{-1/2})$, and hence the result follows since $P_0\{\phi_{\psi}(\bm{Z})\} = 0$. 

\subsection{The remainder term}

Evaluating the remainder $R_n = \E\{\hat{\phi}_s(\bm{Z}) + \hat{\Theta}_s^0  - \Theta_s \} $ gives
\begin{align*}
R_n &= \E\left[ \{\hat{\varphi}-\hat{\tau}_s\}^2 - \{\hat{\varphi}-\hat{\tau}\}^2 - \{\tau-\tau_s\}^2  \right]
\end{align*}
where we have used the fact that $\Theta_s = \E[\{\tau-\tau_s\}^2]$. By algebraic manipulation, we write
\begin{align*}
R_n &= \E\left[ \{\hat{\tau}-\hat{\tau}_s\}^2 - \{\tau-\tau_s\}^2  + 2 \{\hat{\tau}-\hat{\tau}_s\}\{\hat{\varphi}-\hat{\tau}\} \right]
\end{align*}
We then use the identity,
\begin{align*}
\E\left[ \{\hat{\tau}-\hat{\tau}_s\}^2 - \{\tau-\tau_s\}^2\right] &= \E\left[ \{\tau_s-\hat{\tau}_s\}^2-\{\tau-\hat{\tau}\}^2 + 2\{\hat{\tau}-\hat{\tau}_s\}\{\hat{\tau}-\tau\} \right]
\end{align*}
to rewrite the remainder term as the sum of two error terms,
\begin{align*}
R_n &= \E\left[ \{\hat{\tau}-\hat{\tau}_s\}^2 - \{\tau-\tau_s\}^2  + 2 \{\hat{\tau}-\hat{\tau}_s\}\{\hat{\varphi}-\hat{\tau}\} \right] \\
&= \E\left[ \{\tau_s-\hat{\tau}_s\}^2-\{\tau-\hat{\tau}\}^2  + 2 \{\hat{\tau}-\hat{\tau}_s\}\{\hat{\tau} - \tau\} + 2 \{\hat{\tau}-\hat{\tau}_s\}\{\hat{\varphi}-\hat{\tau}\}  \right] \\
  &= \E\left[ \{\tau_s-\hat{\tau}_s\}^2-\{\tau-\hat{\tau}\}^2 + 2 \{\hat{\tau}-\hat{\tau}_s\}\{\hat{\varphi} -\tau\}\right] \\
  &= \underbrace{\E\left[ \{\tau_s-\hat{\tau}_s\}^2-\{\tau-\hat{\tau}\}^2\right]}_{\text{CATE error}} + \underbrace{2\E\left[\{\hat{\tau}-\hat{\tau}_s\} r \right]}_{\text{Pseudo-outcome error}}
\end{align*}
where $r=r(\bm{X})$ is defined by $r(\bm{x}) \equiv \E\left[ \hat{\varphi} | \bm{X}=\bm{x}  \right] - \tau(\bm{x})$, which represents a pseudo-outcome error in the sense that $r(\bm{x})=\E\left[ \hat{\varphi} -\varphi| \bm{X}=\bm{x}  \right]$. Splitting the remainder in to two error terms allows us to consider that the CATE error is $o_P(n^{-1/2})$ when (A2) holds. For the pseudo-outcome error we use the Cauchy-Schwarz inequality to show that
\begin{align*}
\E\left[\{\hat{\tau}-\hat{\tau}_s\} r \right]^2 &\leq \E\left[\{\hat{\tau}-\hat{\tau}_s\}^2 \right] \E\left[r^2 \right] \leq \delta^2 \E\left[r^2 \right]
\end{align*}
Hence the pseudo-outcome error term is $o_P(n^{-1/2})$ if $r$ is $o_P(n^{-1/2})$. By iterated expectation
\begin{align*}
r(\bm{x}) &= \left\{\frac{\pi(\bm{x})}{\hat{\pi}(\bm{x})} - 1 \right\}\left\{\mu(1,\bm{x}) - \hat{\mu}(1,\bm{x}) \right\} - \left\{\frac{1-\pi(\bm{x})}{1-\hat{\pi}(\bm{x})} - 1 \right\}\left\{\mu(0,\bm{x}) - \hat{\mu}(0,\bm{x}) \right\}
\end{align*}
Using the inequality $(a+b)^2 \leq 2(a^2 + b^2)$ then
\begin{align*}
r^2(\bm{x}) &\leq 2\left\{\frac{\pi(\bm{x})}{\hat{\pi}(\bm{x})} - 1 \right\}^2\left\{\mu(1,\bm{x}) - \hat{\mu}(1,\bm{x}) \right\}^2 +2 \left\{\frac{1-\pi(\bm{x})}{1-\hat{\pi}(\bm{x})} - 1 \right\}^2\left\{\mu(0,\bm{x}) - \hat{\mu}(0,\bm{x}) \right\}^2 \\
&\leq \left(\frac{2}{\epsilon^2}\right) \left\{ \pi(\bm{x}) - \hat{\pi}(\bm{x}) \right\}^2 \left[ \left\{\mu(1,\bm{x}) - \hat{\mu}(1,\bm{x}) \right\}^2 + \left\{\mu(0,\bm{x}) - \hat{\mu}(0,\bm{x}) \right\}^2\right]
\end{align*}
with the second inequality follows since $\hat{\pi} \in (\epsilon, 1-\epsilon)$. The final expression above is $o_P(n^{-1})$ under (A1), which completes the proof that $R_n$ itself is $o_P(n^{-1/2})$.

\subsection{The empirical process term}

First write the empirical process term as the sum
\begin{align*}
    H_n &= (\mathbb{P}_n - P_0)\{\Theta_s - \hat{\Theta}_s^0\} \\
    &+ 2 (\mathbb{P}_n - P_0)\{(\hat{\varphi} - \varphi)(\hat{\tau} - \hat{\tau}_s)\} \\
    &+ (\mathbb{P}_n - P_0)\{(\varphi - \hat{\tau}_s)^2 - (\varphi - \tau_s)^2\} \\
    &- (\mathbb{P}_n - P_0)\{(\varphi - \hat{\tau})^2 - (\varphi - \tau)^2\}
\end{align*}
Note that the first term is zero since $(\mathbb{P}_n - P_0)\{\Theta_s - \hat{\Theta}_s^0\} = (\Theta_s - \hat{\Theta}_s^0)(\mathbb{P}_n - P_0)\{1\} = 0$. When the Donsker condition holds, then, by Lemma 19.24 of
%\cite{Vaart1998}
, the second term is $o_P(n^{-1/2})$ provided (i) that $\E\{(\hat{\varphi} - \varphi)^2(\hat{\tau} - \hat{\tau}_s)^2\} = o_p(1)$, the third term is $o_P(n^{-1/2})$ provided (ii) that $\E\left[\left\{(\varphi - \hat{\tau}_s)^2 - (\varphi - \tau_s)^2\right\}^2\right] = o_p(1)$, and the fourth term is $o_P(n^{-1/2})$ provided (iii) that $\E\left[\left\{(\varphi - \hat{\tau})^2 - (\varphi - \tau)^2\right\}^2\right] = o_p(1)$. Similarly, under sample splitting then by Chebyshev's inequality, (i), (ii), and (iii) are also sufficient conditions for $H_n$ to be $o_P(n^{-1/2})$. We will examine these conditions in reverse order.

For (iii) we write
\begin{align*}
    (\varphi - \hat{\tau})^2 - (\varphi - \tau)^2 &= 2(\varphi - \tau)(\tau-\hat{\tau}) + (\tau-\hat{\tau})^2 \\
    \E\left[\left\{(\varphi - \hat{\tau})^2 - (\varphi - \tau)^2\right\}^2|\bm{X}\right] &= 4\Var(\varphi|\bm{X})(\tau-\hat{\tau})^2 + (\tau-\hat{\tau})^4
\end{align*}
Since $\hat{\tau}$ is consistent, and $\Var(\varphi|\bm{X}) < K$ then (iii) holds. Also since $\hat{\tau}_s$ is consistent then (ii) holds in the same way. For (i) we apply Hölder's inequality to obtain
\begin{align*}
    \E\{(\hat{\varphi} - \varphi)^2(\hat{\tau} - \hat{\tau}_s)^2\} &\leq \E\{(\hat{\tau} - \hat{\tau}_s)^2\} ||(\hat{\varphi} - \varphi)^2||_\infty \\
    &\leq \delta^2 ||(\hat{\varphi} - \varphi)^2||_\infty 
\end{align*}
where $||.||_\infty $ denotes the supremum norm. Since $\hat{\varphi}$ is a uniformly consistent estimator of $\varphi$, $||(\hat{\varphi} - \varphi)^2||_\infty = o_p(1)$, thus $H_n = o_P(n^{-1/2})$ which completes the proof.

% and we remark that $\E\{(\hat{\varphi} - \varphi)^2\}$ is the empirical process term which appears in analogous derivations for average treatment effect estimators. Therefore, in view of Theorem 5.1 of \cite{Chernozhukov2018}, this term also converges to zero when (A1) holds, $\hat{\pi} \in (\epsilon, 1-\epsilon)$, and $\Var(Y|\bm{X}) < K$.

\subsection{Proof of Theorem \ref{asym_theorem2}}

Suppose we have two regular asymptotically linear estimators
\begin{align}
\hat{\Theta}_s - \Theta_s &= \mathbb{P}_n\{\phi_s(\bm{Z})\} + o_p(n^{-1/2}) \label{ral_s} \\
\hat{\Theta}_p - \Theta_p &= \mathbb{P}_n\{ \phi_p(\bm{Z}) \}+ o_p(n^{-1/2}) \label{ral_p}
\end{align}
It follows by algebraic manipulations that
\begin{align*}
n^{1/2}(\hat{\Psi}_s - \Psi_s) &= \frac{\Theta_p}{\hat{\Theta}_p} \left[ n^{1/2}\mathbb{P}_n\{ \Phi_s(\bm{Z}) \}+ o_p(1) \right]
\end{align*}
where $\Phi_s(z) = \{\phi_s(z) - \Psi_s\phi_p(z)\}/\Theta_p$ is the IC of $\Psi_s$. By Slutsky's Theorem and the fact that $\hat{\Theta}_p/\Theta_p$ converges to 1 in probability
\begin{align*}
\lim_{n\to \infty} n^{1/2}(\hat{\Psi}_s - \Psi_s) &= \lim_{n\to \infty} n^{1/2}\mathbb{P}_n\{\Phi_s(\bm{Z})\}
\end{align*}
which gives the desired result due to the central limit theorem. We note that this set up is quite general when one considers estimands which are written as the ratio of two other estimands, such as $\Psi_s$ in the present context.

Clearly \eqref{ral_s} follows by Theorem \ref{asym_theorem}. We must therefore check that \eqref{ral_p} also holds. 

Most of the steps in the Proof of Theorem \ref{asym_theorem} can be applied directly to $\Theta_p$. When decomposing the empirical process term, however, we are left with the term
\begin{align*}
    (\mathbb{P}_n - P_0)\{(\varphi - \hat{\tau}_p^*)^2 - (\varphi - \tau_p)^2\}
\end{align*}
in place of the corresponding term involving $\tau_s$. Since $\tau_p$ and $\hat{\tau}_p^*$ are constant this term reduces to
\begin{align*}
    - 2 (\hat{\tau}_p^* - \tau_p)(\mathbb{P}_n - P_0)\{\varphi\}
\end{align*}
Next we note that the conditions of Theorem \ref{asym_theorem} imply that the AIPW is a RAL estimator of the ATE
\begin{align*}
    \hat{\tau}_p^* - \tau_p = \mathbb{P}_n\{\varphi - \tau_p\} + o_P(n^{-1/2})
\end{align*}
See e.g. Theorem 5.1 of \cite{Chernozhukov2018}. Also, $(\mathbb{P}_n - P_0)\{\varphi\} \overset{p}{\to} 0$ by the weak law of large numbers. Therefore this term in the empirical process term decomposition will be $o_P(n^{-1/2})$, which completes the proof.

\section{Additional plots for simulation results}
\label{appendix:additional_plots}
\begin{figure}[htbp]
\centering
\includegraphics[width=\linewidth]{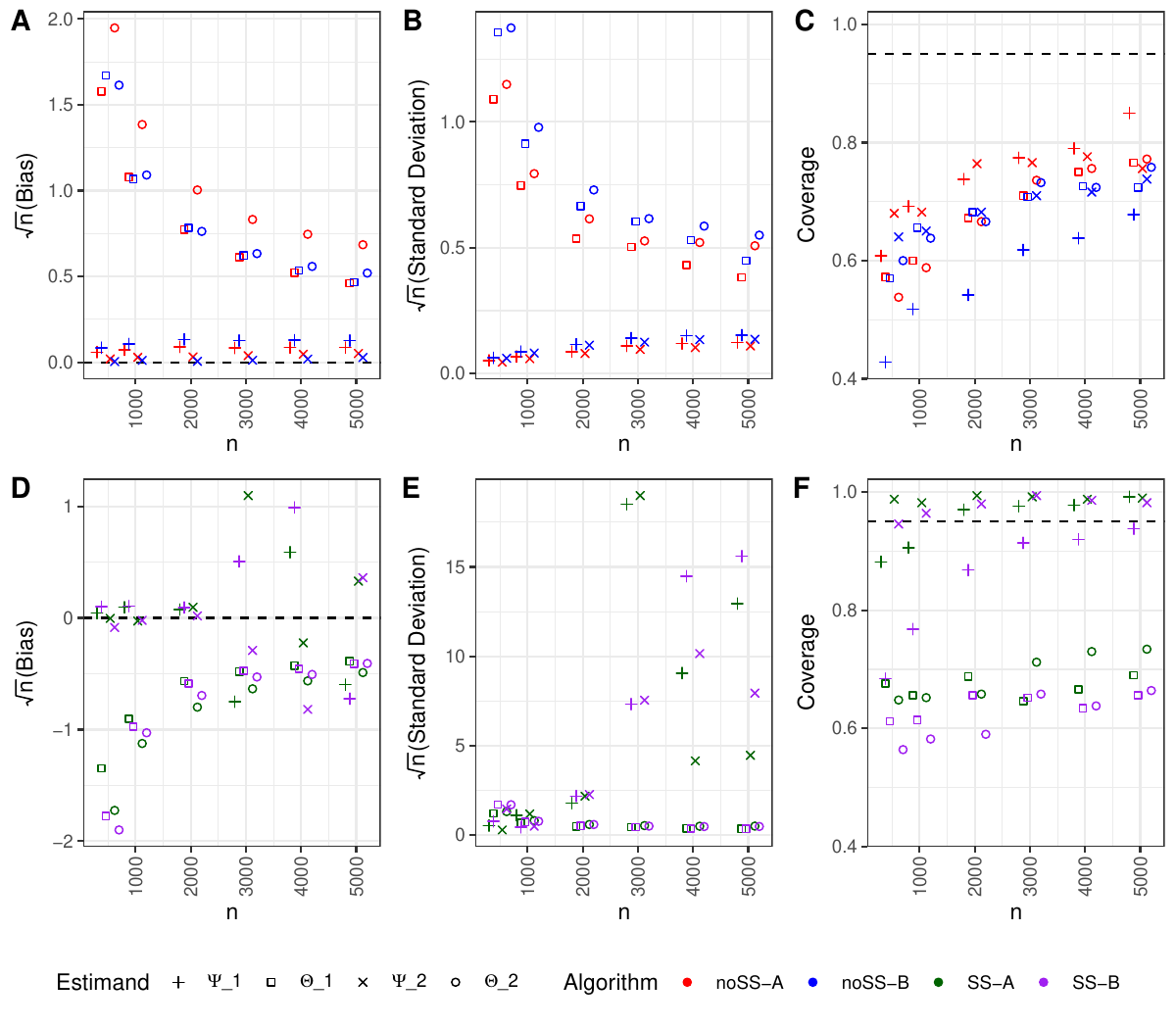}
\caption{Bias (A, D), empirical standard deviation (B, E) and coverage (C, F) for estimators from DGP 2. Dashed lines indicate zero bias, and nominal 95\% CI coverage. For readability, a small amount of `jitter' has been added to the sample size, $n$. In order to present scaled and unscaled TE-VIMs together, the standard deviation and bias of scaled TE-VIMs has been multiplied by the true VTE. Note that the bias and variance scales differ between sub-plots A,B and D,E.}
\label{sim_results_dgp2}
\end{figure}

\begin{figure}[htbp]
\centering
\includegraphics[width=\linewidth]{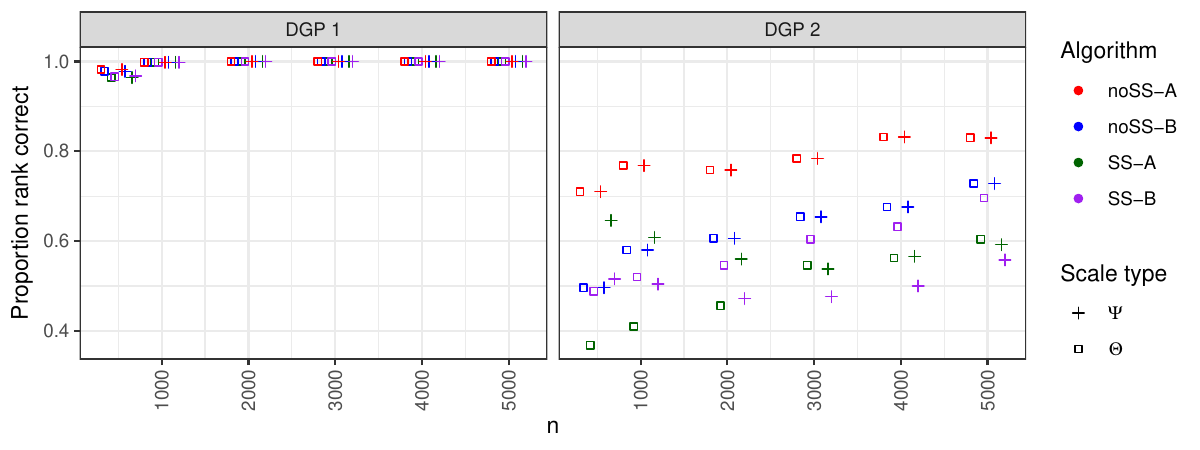}
\caption{Empirical probability that (scaled) TE-VIMs recover the correct importance ranking for DGPs 1 and 2. For readability, a small amount of `jitter' has been added to $n$.}
\label{sim_results_ranking}
\end{figure}

\begin{figure}[htbp]
\centering
\includegraphics[width=\linewidth]{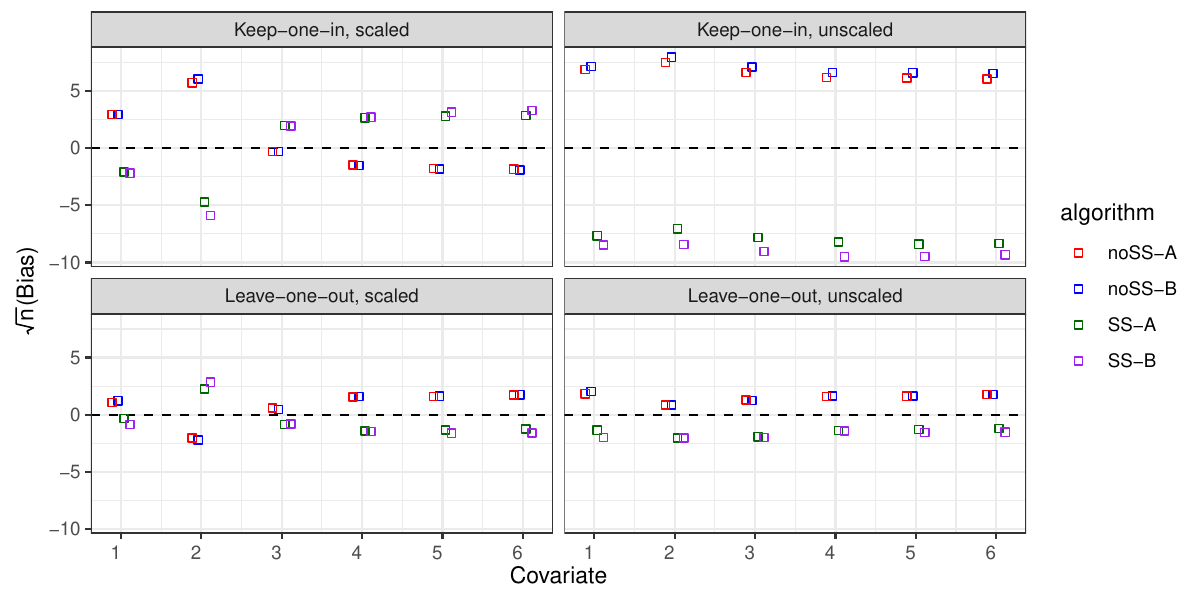}
\caption{Empirical bias for estimators from DGP 3. Dashed lines indicate zero bias. In order to present scaled and unscaled TE-VIMs together, the bias of scaled TE-VIMs has been multiplied by the true VTE $\Theta_p = 8$. Here scaled and unscaled refers to estimators of the type $\Psi$ and $\Theta$ respectively.}
\label{sim_results_dgp3_bias}
\end{figure}

\begin{figure}[htbp]
\centering
\includegraphics[width=\linewidth]{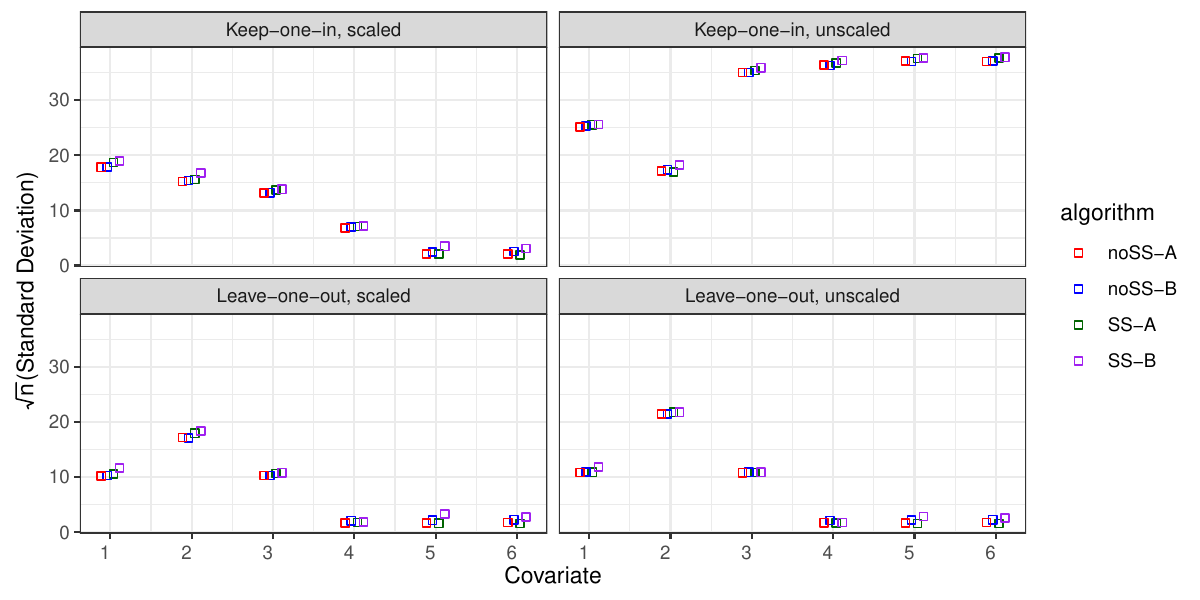}
\caption{Empirical standard deviation of estimators from DGP 3. In order to present scaled and unscaled TE-VIMs together, the standard deviation of scaled TE-VIMs has been multiplied by the true VTE $\Theta_p = 8$. Here scaled and unscaled refers to estimators of the type $\Psi$ and $\Theta$ respectively.}
\label{sim_results_dgp3_variance}
\end{figure}

\begin{figure}[htbp]
\centering
\includegraphics[width=\linewidth]{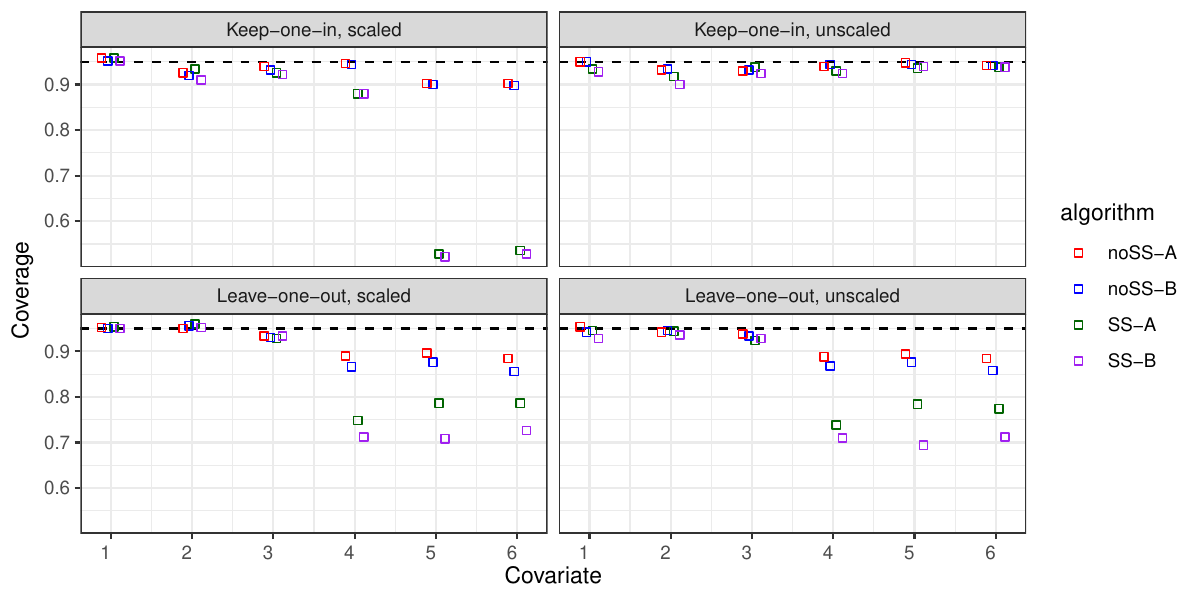}
\caption{Empirical CI coverage for estimators from DGP 3. Dashed lines indicate nominal 95\% CI coverage. Here scaled and unscaled refers to estimators of the type $\Psi$ and $\Theta$ respectively.}
\label{sim_results_dgp3_coverage}
\end{figure}

\clearpage
\section{Additional discussion}
\label{appendix:additional_discussion}

\subsection{Representations of TE-VIMs}
\label{supp:representations}

We claim that the following representations of $\Theta_s$ are equivalent
\begin{align*}
    \Theta_s &= \E[\{\tau(\bm{X}) - \tau_s(\bm{X})\}^2] \\
    &= \E[\Var\{\tau(\bm{X})|\bm{X}_{-s}\}]\\
    &= \Var\{\tau(\bm{X})\} - \Var\{\tau_s(\bm{X})\}
\end{align*}
Proof: The second line follows from the first by noting that $\tau_s(\bm{x}) = \E\{\tau(\bm{X})|\bm{X}_{-s} = \bm{x}_{-s}\}$. The third line follows from the first since
\begin{align*}
    \Theta_s &= \E[\{\tau(\bm{X}) - \tau_s(\bm{X})\}^2] \\
    &= \E\{\tau^2(\bm{X})\} + \E\{\tau_s^2(\bm{X})\} - 2 \E\{\tau(\bm{X})\tau_s(\bm{X})\} \\
    &= \E\{\tau^2(\bm{X})\} - \E\{\tau_s^2(\bm{X})\} \\
    &= \left[\E\{\tau^2(\bm{X})\} - \tau_p^2\right] - \left[\E\{\tau_s^2(\bm{X})\} - \tau_p^2\right] \\
    &= \Var\{\tau(\bm{X})\} - \Var\{\tau_s(\bm{X})\}
\end{align*}
where we the law of iterated expectations to obtain $\E\{\tau(\bm{X})\tau_s(\bm{X})\} = \E\{\E\{\tau(\bm{X})\tau_s(\bm{X})|\bm{X}_{-s}\}\} = \E\{\tau^2_s(\bm{X})\}$.

\subsection{Invariance to transformations}
\label{supp:invariance}

The scaled TE-VIM $\Psi_s$ is invariant to linear outcome transformations. To see why, let $\tilde{Y} = b Y + k$ and $\tilde{Y}^a = b Y^a + k$ for constants $b \neq 0$ and $k$. Letting superscript tilde to denote the modified values,
\begin{align*}
\tilde{\tau}(x) &\equiv \E(\tilde{Y}^1 - \tilde{Y}^0 | \bm{X}=\bm{x}) = b\tau(\bm{x}) \\
\implies \tilde{\Theta}_s &\equiv \E[\Var\{\tilde{\tau}(\bm{X})|\bm{X}_{-s}\}] = b^2 \Theta_s \\
\implies \tilde{\Psi}_s &\equiv\frac{\tilde{\Theta}_s}{\tilde{\Theta}_p} = \frac{\Theta_s}{\Theta_p} = \Psi_s.
\end{align*}
Moreover, $\Psi_s$ is invariant to invertible component wise mappings of $\bm{X}$. To see why, consider a mapping $g:\mathbb{R}^p\mapsto\mathbb{R}^p$ such that 
\begin{align*}
    g(\bm{x}) = (g_1(x_1), ..., g_p(x_p))
\end{align*}
where $g_j:\mathbb{R}\mapsto\mathbb{R}$ is an invertible function for $j \in \{1,...,p\}$. The claim follows since the conditional distribution of $Y$ induced by $\bm{X}_{-s}$ is the same as that induced by $g(\bm{X})_{-s}$. In particular, using superscript tilde to denote a different set of modified values,
\begin{align*}
\tilde{\tau}(g(\bm{x})) &\equiv \E(Y^1 - Y^0 | g(\bm{X})=g(\bm{x})) = \tau(\bm{x}) \\
\implies \tilde{\Theta}_s &\equiv \E[\Var\{\tilde{\tau}(g(\bm{X}))|g(\bm{X})_{-s}\}] = \E[\Var\{\tau(\bm{X})|g(\bm{X})_{-s}\}] = \Theta_s \\
\implies \tilde{\Psi}_s &\equiv\frac{\tilde{\Theta}_s}{\tilde{\Theta}_p} = \frac{\Theta_s}{\Theta_p} = \Psi_s.
\end{align*}

% \subsection{Example of estimator incompatibility}

% Suppose that for $p=2$ covariates, the true CATE is linear in $X_1$, i.e. $\tau(x) = \alpha_0 + \alpha_1 x_1$, with $\tau_2(x) = \tau(x)$ and $\tau_1(x) = \alpha_0 + \alpha_1 \E(X_1|X_2 = x_2)$. Hence $\Theta_2 = 0$ and $\Theta_1 = \alpha_1^2 \E\{\Var(X_1|X_2)\}$, and thus $\Psi_2 = 0$ and $\Psi_1 = \E\{\Var(X_1|X_2)\} / \Var(X_1)$.

% Next suppose that $\hat{\tau}$ is estimated using the DR-learner with a lasso penalty as $\hat{\tau}(x) = \hat{\alpha}_0 + \hat{\alpha}_1 x_1 + \hat{\alpha}_2 x_2$. 

\subsection{Shapley values}
\label{supp:shapley}

We use the Shapley value definition given by \cite{Williamson2020} for an arbitrary value/ loss function. For covariate $j \in \{1, ..., p\}$, and letting $\mathcal{P}_j$ denote the power set (set of all possible subsets) of $\{1,...,p\}\setminus\{j\}$, we define the TE-VIM Shapley value
\begin{align*}
    S_j \equiv \sum_{s \in \mathcal{P}_j} \frac{1}{p} \binom{p-1}{|s|}^{-1} \{\Theta_{s \cup \{j\}} - \Theta_{s}\}
\end{align*}
Since $\Theta_{s \cup \{j\}} \geq \Theta_{s}$ for all $s$, then $S_j \geq 0$. Also, by construction
\begin{align*}
    \Theta_p = \sum_{j=1}^p S_j.
\end{align*}

\subsection{Plug-in estimators}
\label{supp:plugin}

Plug-in estimators for the TE-VIM $\Theta_s$ and the scaled TE-VIM $\Psi_s$ are non-trivial to construct using estimators for $\tau(\bm{x})$ and $\tau_s(\bm{x})$. We illustrate this by considering plug-in estimation of the regression-VIM $\Omega_s$ which suffers similar difficulties. Writing the regression-VIM as
\begin{align*}
    \Omega_s = \E\left[\{\mu(\bm{X}) - \mu(_s\bm{X})\}^2\right]
\end{align*}
\cite{Williamson2021} construct a `naive plug-in' estimator
\begin{align*}
    \hat{\Omega}^0_s = n^{-1}\sum_{i=1}^n \{\hat{\mu}(\bm{x}_i) - \hat{\mu}(_s\bm{x}_i)\}^2.
\end{align*}
However, this estimator is not strictly plug-in in the estimand mapping sense discussed in the main paper. To see why, we write $\Omega_s$ in terms of $\mu(\bm{x})$, the distribution of $\bm{X}_{s}$ given $\bm{X}_{-s}$, which we denote by the measure $dP_0(\bm{x}_{s}|\bm{x}_{-s})$, and the distribution of $\bm{X}_{-s}$, denoted $dP_0(\bm{x}_{s})$,
\begin{align*}
    \Omega_s = \int \left[\mu(\bm{x}) - \left\{\int \mu(\bm{x}) \underbrace{dP_0(\bm{x}_{s}|\bm{x}_{-s})}_{\text{regression}}\right\}\right]^2 \underbrace{dP_0(\bm{x}_{s}|\bm{x}_{-s})}_{\text{empirical}}dP_0(\bm{x}_{s}).
\end{align*}
Hence, in the estimand definition, the distribution of $\bm{X}_{s}$ given $\bm{X}_{-s}$ appears twice. However, the $\hat{\Omega}^0_s$ estimator effectively uses, two different (and possibly inconsistent) implicit distributions to approximate $dP_0(\bm{x}_{s}|\bm{x}_{-s})$. The first is implied by the regression estimate $\hat{\mu}_s(\bm{x})$, whilst the second is implied by the empirical distribution of covariates. In particular, this inconsistency means that there is no guarantee that 
\begin{align*}
    n^{-1}\sum_{i=1}^n \hat{\mu}(\bm{x}_i) = n^{-1}\sum_{i=1}^n \hat{\mu}_s(\bm{x}_i)
\end{align*}
without additional steps taken to ensure that this identity holds e.g. through targeting (in a TMLE sense) an initial regression estimator of $\hat{\mu}_s(\bm{x})$.

We remark that for the Proofs in Appendix \ref{Appen_assymp}, the exact form of the initial TE-VIM estimator $\hat{\Theta}^0_s$ does not affect the form of the final estimator $\hat{\Theta}_s$, which, can be thought of as a bias-corrected version of $\hat{\Theta}^0_s$. As such, one could let $\hat{\Theta}^0_s = n^{-1}\sum_{i=1}^n \{\hat{\tau}(\bm{x}_i) - \hat{\tau}(_s\bm{x}_i)\}^2 $, though this estimand is also not plug-in for the reasons above.

\subsection{TE-VIM estimation on different scales}\label{TE-VIMscales}

The TE-VIM $\Theta_s \geq 0$ and the scaled TE-VIM $\Psi_s\in[0,1]$ are both bounded, therefore one might want to perform inference on scales which respect these bounds. For instance, one may prefer to treat $\log(\Theta_s)$ as the target estimand, with the assumption that $\Theta_s > 0$, or treat $\logit(\Psi_s)$ as the target estimand, assuming $\Psi_s\in(0,1)$. Here we sketch how one-step bias correction estimators could be constructed for these alternatives, and derive their asymptotic distributions.

First consider that, since the IC represents a pathwise derivatives for $\log(\Theta_s)$ and $\logit(\Psi_s)$ respectively are
\begin{align*}
&\frac{\phi_s(\bm{Z})}{\Theta_s} \\
&\frac{\Phi_s(\bm{Z})}{\Psi_s(1-\Psi_s)} 
\end{align*}
We let $\hat{\Theta}^0_s$ and $\hat{\Psi}^0_s$ be initial estimators such that $\hat{\Theta}^0_s >0$ and $\hat{\Psi}^0_s \in (0, 1)$. E.g.
\begin{align*}
    \hat{\Theta}^0_s &= n^{-1} \sum_{i=1}^n \{ \hat{\tau}(x_i) - \hat{\tau}_s(x_i)\}^2 \\
    \hat{\Theta}^0_p &= \hat{\Theta}^0_s + n^{-1} \sum_{i=1}^n \left[ \hat{\tau}_s(x_i) - \left\{n^{-1} \sum_{i=1}^n\hat{\tau}_s(x_i) \right\} \right]^2 \\
    \hat{\Psi}^0_s &= \frac{\hat{\Theta}^0_s}{\hat{\Theta}^0_p}
\end{align*}
Using these initial estimators one can construct the one-step bias corrected estimators 
\begin{align*}
\log(\hat{\Theta}^0_s) &+ \frac{n^{-1}\sum_{i=1}^n \hat{\phi}_s(\bm{z}_i)}{\hat{\Theta}^0_s} \\
\logit(\hat{\Psi}^0_s) &+ \frac{n^{-1}\sum_{i=1}^n \hat{\Phi}_s(\bm{z}_i)}{\hat{\Psi}^0_s(1-\hat{\Psi}^0_s)}
\end{align*}
which we rewrite in terms of the estimators in the main text as
\begin{align}
\log(\hat{\Theta}^0_s) &+ \frac{(\hat{\Theta}_s - \hat{\Theta}^0_s)}{\hat{\Theta}^0_s} \label{log_extra} \\
\logit(\hat{\Psi}^0_s) &+ \frac{\hat{\Theta}_p}{\hat{\Theta}_p^0} \frac{(\hat{\Psi}_s - \hat{\Psi}_s^0)}{\hat{\Psi}^0_s(1-\hat{\Psi}^0_s)} \label{logit_extra}
\end{align}
where we have used the fact that
\begin{align*}
n^{-1}\sum_{i=1}^n\hat{\Phi}_s(\bm{z}_i) &= \frac{n^{-1}\sum_{i=1}^n\hat{\phi}_s(\bm{z}_i) - \hat{\Psi}^0_s n^{-1}\sum_{i=1}^n\hat{\phi}_p(\bm{z}_i)}{\hat{\Theta}_p^0} \\
&= \frac{(\hat{\Theta}_s - \hat{\Theta}_s^0) - \hat{\Psi}^0_s(\hat{\Theta}_p - \hat{\Theta}_p^0)}{\hat{\Theta}_p^0} \\
&= \frac{\hat{\Theta}_p}{\hat{\Theta}_p^0}(\hat{\Psi}_s - \hat{\Psi}_s^0)
\end{align*}
We remark that, unlike the estimators $\hat{\Theta}_s$ and $\hat{\Psi}_s$, the estimators in \eqref{log_extra} and \eqref{logit_extra} depend on the initial estimators $\hat{\Theta}^0_s$ and $\hat{\Psi}^0_s$ in a non-trivial way. We derive asymptotic distributions of these estimators given additional conditions on these initial estimators.

\begin{theorem}
    Assume the conditions of Theorem \ref{asym_theorem} hold, $\Theta_s > 0$, and $(\Theta_s - \hat{\Theta}^0_s)/\hat{\Theta}^0_s = o_P(n^{-1/4})$, then the estimator in \eqref{log_extra} converges to $\log(\Theta_s)$ in probability, with a difference that, when multiplied by $n^{1/2}$, converges to a mean-zero normal random variable, with variance $||\phi_s(\bm{Z})||^2/\Theta_s^2$.
\end{theorem}
\begin{proof}
    Under the conditions of Theorem \ref{asym_theorem}, then $\hat{\Theta}_s$ is regular asymptotically linear, i.e. we can write
    \begin{align*}
        \hat{\Theta}_s = \Theta_s + n^{-1}\sum_{i=1}^n\phi_s(\bm{z}_i) + o_P(n^{-1/2})
    \end{align*}
    Hence,
    \begin{align*}
        \log(\hat{\Theta}^0_s) + \frac{(\hat{\Theta}_s - \hat{\Theta}^0_s)}{\hat{\Theta}^0_s} - \log(\Theta_s) &= n^{-1}\sum_{i=1}^n \frac{\phi_s(\bm{z}_i)}{\hat{\Theta}^0_s} + \frac{o_P(n^{-1/2})}{\hat{\Theta}^0_s} + \frac{\Theta_s - \hat{\Theta}^0_s}{\hat{\Theta}^0_s} - \log\left(1 + \frac{\Theta_s - \hat{\Theta}^0_s}{\hat{\Theta}^0_s}\right)
    \end{align*}
    Using the Taylor series of $\log(1+x)$
    \begin{align*}
        &= n^{-1}\sum_{i=1}^n \frac{\phi_s(\bm{z}_i)}{\hat{\Theta}^0_s} + \frac{o_P(n^{-1/2})}{\hat{\Theta}^0_s} + \sum_{j=2}^\infty  \frac{(-1)^{j+1}}{j}\left(\frac{\Theta_s - \hat{\Theta}^0_s}{\hat{\Theta}^0_s}\right)^j \\
        &=\left\{n^{-1}\sum_{i=1}^n \frac{\phi_s(\bm{z}_i)}{\Theta_s} + \frac{o_P(n^{-1/2})}{\Theta_s}\right\}(1 + \hat{u}) + \sum_{j=2}^\infty  \frac{(-1)^{j+1} \hat{u}^j}{j}
    \end{align*}
    where $\hat{u} \equiv (\Theta_s - \hat{\Theta}^0_s)/\hat{\Theta}^0_s = o_P(n^{-1/4})$ thus
    \begin{align*}
        \log(\hat{\Theta}^0_s) + \frac{(\hat{\Theta}_s - \hat{\Theta}^0_s)}{\hat{\Theta}^0_s} - \log(\Theta_s) &=n^{-1}\sum_{i=1}^n \frac{\phi_s(\bm{z}_i)}{\Theta_s} + o_P(n^{-1/2})
    \end{align*}
\end{proof}
\begin{theorem}
    Assume the conditions of Theorem \ref{asym_theorem2} hold, $\Psi_s \in (0,1)$, $(\Psi_s - \hat{\Psi}^0_s)/\{\hat{\Psi}^0_s(1-\hat{\Psi}^0_s)\} = o_P(n^{-1/4})$, and $(\Theta_p - \hat{\Theta}_p^0)/\hat{\Theta}_p^0 = o_P(n^{-1/4})$ then the estimator in \eqref{logit_extra} converges to $\logit(\Psi_s)$ in probability, with a difference that, when multiplied by $n^{1/2}$, converges to a mean-zero normal random variable, with variance $||\Phi_s(\bm{Z})||^2/\{\Psi_s(1-\Psi_s)\}^2$.
\end{theorem}
\begin{proof}
    Under the conditions of Theorem \ref{asym_theorem2}, then $\hat{\Psi}_s$ is regular asymptotically linear, i.e. we can write
    \begin{align*}
        \hat{\Psi}_s = \Psi_s + n^{-1}\sum_{i=1}^n\Phi_s(\bm{z}_i) + o_P(n^{-1/2})
    \end{align*}
    Also, using the Taylor series of $\log(1\pm x)$ note that for arbitrary values $a,b$
    \begin{align*}
        \logit(a) - \logit(b) &= \log\left(1 + \frac{a-b}{b}\right) - \log\left(1 - \frac{a-b}{1-b}\right) \\
        &= \sum_{j=1}^\infty \frac{(-1)^{j+1}}{j}\left(\frac{a-b}{b}\right)^j + \sum_{j=1}^\infty \frac{1}{j}\left(\frac{a-b}{1-b}\right)^j \\
        &=  \sum_{j=1}^\infty \left(\frac{a-b}{b(1-b)}\right)^j\frac{(b^j - (b-1)^j)}{j}
    \end{align*} 
    Hence,
    \begin{align*}
        &\logit(\hat{\Psi}^0_s) + \frac{\hat{\Theta}_p}{\hat{\Theta}_p^0} \frac{(\hat{\Psi}_s - \hat{\Psi}_s^0)}{\hat{\Psi}^0_s(1-\hat{\Psi}^0_s)} - \logit(\Psi_s) \\
        &= \left[n^{-1}\sum_{i=1}^n\frac{\Phi_s(\bm{z}_i)}{\hat{\Psi}^0_s(1-\hat{\Psi}^0_s)} +   \frac{(\Psi_s - \hat{\Psi}_s^0)}{\hat{\Psi}^0_s(1-\hat{\Psi}^0_s)} + \frac{o_P(n^{-1/2})}{\hat{\Psi}^0_s(1-\hat{\Psi}^0_s)}\right] \frac{\hat{\Theta}_p}{\hat{\Theta}_p^0}\\
        &-\sum_{j=1}^\infty  \left(\frac{\Psi_s - \hat{\Psi}^0_s}{\hat{\Psi}^0_s(1-\hat{\Psi}^0_s)}\right)^j\frac{((\hat{\Psi}^0_s)^j - (\hat{\Psi}^0_s-1)^j)}{j} \\
        &= \left[n^{-1}\sum_{i=1}^n\frac{\Phi_s(\bm{z}_i)}{\Psi_s(1-\Psi_s)}  + \frac{o_P(n^{-1/2})}{\Psi_s(1-\Psi_s)}\right]\frac{\hat{\Theta}_p}{\hat{\Theta}_p^0} \left( 1 + \frac{\Psi_s(1-\Psi_s) -\hat{\Psi}^0_s(1-\hat{\Psi}^0_s)}{\hat{\Psi}^0_s(1-\hat{\Psi}^0_s)}\right) \\
        &+ \hat{v}\hat{u} -\sum_{j=2}^\infty  \hat{u}^j \frac{((\hat{\Psi}^0_s)^j - (\hat{\Psi}^0_s-1)^j)}{j}
    \end{align*}
    where $\hat{u} \equiv (\Psi_s - \hat{\Psi}^0_s)/\{\hat{\Psi}^0_s(1-\hat{\Psi}^0_s)\} = o_P(n^{-1/4})$ and
    \begin{align*}
    \hat{v} &= \frac{\hat{\Theta}_p}{\hat{\Theta}_p^0} - 1\\
    &= \frac{\hat{\Theta}_p - \Theta_p}{\hat{\Theta}_p^0} + \frac{\Theta_p - \hat{\Theta}_p^0}{\hat{\Theta}_p^0} = o_p(n^{-1/4})
    \end{align*}
    where the last line  follows since $\hat{\Theta}_p - \Theta_p = o_p(n^{-1/4})$. Also
    \begin{align*}
        \frac{\hat{\Theta}_p}{\hat{\Theta}_p^0} \left\{ 1 + \frac{\Psi_s(1-\Psi_s) -\hat{\Psi}^0_s(1-\hat{\Psi}^0_s)}{\hat{\Psi}^0_s(1-\hat{\Psi}^0_s)}\right\} = (1 + \hat{v}) \left\{ 1 + \hat{u}(1 - \Psi_s - \hat{\Psi}^0_s)\right\}\overset{p}{\to} 1
    \end{align*}
    Therefore we recover
        \begin{align*}
        \logit(\hat{\Psi}^0_s) + \frac{\hat{\Theta}_p}{\hat{\Theta}_p^0} \frac{(\hat{\Psi}_s - \hat{\Psi}_s^0)}{\hat{\Psi}^0_s(1-\hat{\Psi}^0_s)} - \logit(\Psi_s) &= n^{-1}\sum_{i=1}^n\frac{\Phi_s(\bm{z}_i)}{\Psi_s(1-\Psi_s)}  + o_P(n^{-1/2}).
    \end{align*}
\end{proof}

\subsection{Defining Treatment effects on different scales}
\label{supp:logte}

In the current work, we examine the importance of variable subsets in predicting the causal contrast $Y^1 - Y^0$, and hence the CATE $\tau(\bm{x})$. Our conclusions regarding heterogeneity depend on this choice of scale, and different conclusions could be reached if one had considered another effect definition. For instance, supposing that $Y>0$ almost surely, then one might be interested in VIMs with respect to the conditional risk ratio
\begin{align}
    \psi(\bm{x}) \equiv \log\{\E(Y^1|\bm{X}=\bm{x})\} - \log\{\E(Y^0|\bm{X}=\bm{x})\}. \label{cond_log}
\end{align}
It is possible that $\psi(\bm{x})$ is constant (suggesting no heterogeneity), but $\tau(\bm{x})$ is not constant (suggesting some heterogeneity), or vice-versa. Similar problems apply to conditional odds ratios, where for a binary outcome $Y\in\{0,1\}$, one replaces the logarithms in \eqref{cond_log}, with the logit function. It is an open topic of debate, how treatment effects should be communicated to clinicians in such settings %\citep{Huitfeldt2021, Shannin2022}.
We recommend that practitioners remain aware of scale dependencies when using TE-VIMs.

\subsection{Linear projections of the CATE}

The ideas in the current paper have been extended in the direction of nonparametric linear projection parameters by \cite{Boileau2022}
. They propose using the estimands
\begin{align*}
    \beta_j &\equiv \argmin_{\beta \in \mathbb{R}} L\{\tau_p + \beta [X_j - \E(X_j)]\} = \frac{\Cov(Y^1-Y^0, X_j)}{\Var(X_j)}
\end{align*}
as a proxy for the importance of a covariate $X_j$. Since these estimands are not invariant to the scale on which $X_j$ is defined, hence the authors determine relative variable importance based on the null hypothesis tests that each $\beta_j = 0$. These estimands are generally less sensitive to non-linearities and parameter interactions than TE-VIMs. For instance, it may be the case that $X_j$ is important in explaining treatment effect heterogeneity, but $\beta_j = 0$ in truth. Inference of $\beta_j$ has no power to detect such covariates. That said, the linear term $\beta_j$ remains scientifically interesting, e.g. in roughly determining covariate thresholds for further investigation.

\subsection{Treatment effect cumulative distribution function}
\label{supp:tecdf}

A related proposal considers the treatment effect cumulative distribution function (TE-CDF) \citep{Levy2018}
, which is a curve $\beta: \mathbb{R} \mapsto [0,1]$, with $\beta(t) = Pr\{\tau(\bm{X})\leq t\}$.

Motivated by OTRs, the value $\beta(0)$ is of particular interest since it captures the marginal probability that an individual has a negative CATE, and therefore the proportion of the population which is not treated under the OTR. We note that $\beta(0)$ is not the same as $Pr(Y^1 - Y^0 \leq 0)$ which suffers similar identifiability issues regarding the joint distribution of $(Y^1,Y^0)$ as the quantity $\Var(Y^1-Y^0)$ mentioned previously. Like the OTR-VIMs above, the TE-CDF is generally not pathwise differentiable, hence \cite{Levy2018}
focus instead on a kernel smoothed analogue of $\beta(t)$. It is mentioned by \cite{Levy2021} that, provided $\tau_p>0$, then Chebyshev's inequality implies $\beta(0) \leq \Lambda \equiv \Theta_p / \tau_p^2$.

Thus, the VTE is also of scientific interest since it can be used to bound $\beta(0)$, informing investigators about the probability of negative CATEs once a positive ATE has been established. Estimation of $\Lambda$ could be carried out using estimating equations estimators, as in the current work, or targeted methods \citep{Levy2021}, using the IC for $\Lambda$,
\begin{align*}
\frac{\{\varphi(\bm{Z})-\tau_p\}^2 - \{\varphi(\bm{Z})-\tau(\bm{X})\}^2 - \Lambda \tau_p\{2\varphi(\bm{Z})-\tau_p\} }{\tau_p^2}.
\end{align*}
Below we briefly sketch the details for the estimating equations estimator.

First note Chebyshev's inequality: For a variable $V$ with mean $\mu$ and variance $\sigma^2$, for $k>0$
\begin{align*}
Pr (|V-\mu|\geq k\sigma) &\leq k^{-2} \\
Pr (V \geq \mu + k\sigma) + Pr (V \leq \mu - k\sigma) &\leq k^{-2}
\end{align*}
which implies the weaker inequality,
\begin{align*}
Pr (V \leq \mu - k\sigma) &\leq k^{-2}
\end{align*}
Let $\tau(\bm{X})$ be the CATE with ATE $\tau_p$ and VTE $\Theta_p$ then,
\begin{align*}
\beta(0) = Pr \left\{\tau(\bm{X}) \leq 0\right\} &= Pr \left\{\tau(\bm{X}) \leq \tau_p - \left(\frac{\tau_p}{\sqrt{\Theta_p}}\right)\sqrt{\Theta_p} \right\} \leq \frac{\Theta_p}{\tau^2_p}
\end{align*}
Where the inequality applies only when $\tau_p>0$. It follows that, when the ATE is positive, the quantity on the RHS bounds $\beta(0)$ from above. The quotient rule gives that the IC (pathwise derivative) is,
\begin{align*}
\phi_\beta(\bm{Z}) &= \frac{1}{\tau_p^2}\phi_p(\bm{Z}) -2\left(\frac{\Theta_p}{\tau^3_p}\right)\{\varphi(\bm{Z})-\tau_p\} \\
&= \frac{\{\varphi(\bm{Z})-\tau_p\}^2 - \{\varphi(\bm{Z})-\tau(\bm{X})\}^2 - \left(\frac{\Theta_p}{\tau^2_p}\right) \tau_p\{2\varphi(\bm{Z})-\tau_p\} }{\tau_p^2}
\end{align*}
where $\{\varphi(\bm{Z})-\tau_p\}$ is the IC of $\tau_p$. An estimating equations estimator is that which solves
\begin{align*}
n^{-1}\sum_{i=1}^n \hat{\phi}_\beta(\bm{z}_i) &= 0
\end{align*}
where $\hat{\phi}_\beta(\bm{z})$ is an estimate of $\phi_\beta(\bm{z})$. Therefore $\hat{\Theta}_p/\hat{\tau}_p^2$ is an estimating equations estimator where $\hat{\Theta}_p$ is the VTE estimator in the current paper and $\hat{\tau}_p$ is the AIPW estimator of the ATE.

\subsection{Continuous analogue estimands}
\label{supp:continuo}

Let $\lambda(\bm{x}) \equiv \Cov(A,Y|\bm{X}=\bm{x})/\Var(A|\bm{X}=\bm{x})$.
Consider the loss $L_{P_0}\{f\} \equiv ||\lambda(\bm{x}) - f(\bm{x})||^2$. Applying the same approach as in Appendix \ref{ic_deriv_append}, we see that this loss has IC
\begin{align}
    \{\lambda(\tilde{\bm{x}}) - f(\tilde{\bm{x}})\}^2 - L_{P_0}\{f\} + 2\E[\{\lambda(\bm{x}) - f(\bm{x})\}\partial_t \lambda_{t}(\bm{x})]
\end{align}
where $\lambda_{t}(\bm{x})$ denotes $\lambda(\bm{x})$ evaluated under $P_t$. We will show that,
\begin{align}
\partial_t \lambda(\bm{x}) &= \frac{\tilde{f}(\bm{x})}{f(\bm{x})}\{\tilde{y}-\mu(\bm{x}) - \lambda(\bm{x})\{\tilde{a}-\pi(\bm{x})\}\}\frac{\tilde{a} - \pi(\bm{x})}{\Var(A|\bm{X}=\bm{x})} \label{pseudo_est2}
\end{align}
and hence, letting
\begin{align*}
\varphi_\lambda(\bm{z}) \equiv \{y-\mu(\bm{x}) - \lambda(\bm{x})\{a-\pi(\bm{x})\}\}\frac{a - \pi(\bm{x})}{\Var(A|\bm{X}=\bm{x})} + \lambda(\bm{x})
\end{align*}
then the IC of $L_{P_0}\{f\}$ is
\begin{align}
    \{\varphi_\lambda(\bm{z}) - f(\tilde{\bm{x}})\}^2 - \{\varphi_\lambda(\bm{z}) - \lambda(\tilde{\bm{x}})\}^2- L_{P_0}\{f\}.
\end{align}
Just as with the $L^*\{f\}$ loss in the main paper, this implies the IC of $\E\left[\Var\{\lambda(\bm{X})|\bm{X}_{-s}\}\right]$ is,
\begin{align*}
\{\varphi_\lambda(\bm{z}) - \lambda_s(\bm{x})\} - \{\varphi_\lambda(\bm{z}) - \lambda(\bm{x})\} - \E\left[\Var\{\lambda(\bm{X})|\bm{X}_{-s}\}\right]
\end{align*}
where $\lambda_s(\bm{x})=\E\{\lambda(\bm{X})|\bm{X}_{-s}=\bm{x}_{-s}\}$. The IC for $\Var\{\lambda(\bm{X})\}$ follows as a special case where $s$ includes all the observed covariates. Additionally, by \eqref{quick_lemma2} the IC of $\E\{\lambda(\bm{X})\}$ is,
\begin{align*}
\varphi_\lambda(\bm{z}) - \E\{\lambda(\bm{X})\}.
\end{align*}
To demonstrate \eqref{pseudo_est2} we first note that, by \eqref{quick_lemma},
\begin{align*}
\partial_t \Cov_{P_t}(A,Y|\bm{X}=\bm{x}) &= \partial_t \E_{P_t}\{[A-\E_{P_t}(A|\bm{X})][Y-\E_{P_t}(Y|\bm{X})]|\bm{X}=\bm{x}\} \\
&=  \frac{\tilde{f}(\bm{x})}{f(\bm{x})} \left[\{\tilde{a}-\pi(\bm{x})\}\{\tilde{y}-\mu(\bm{x})\} - \Cov_{P}(A,Y|\bm{X}=\bm{x}) \right]
\end{align*}
We also obtain $\partial_t \Var_{P_t}(A|\bm{X}=\bm{x})$ as a special case of the above expression when $Y=A$. By the quotient rule,
\begin{align*}
\partial_t \frac{\Cov_{P_t}(A,Y|\bm{X}=\bm{x})}{\Var_{P_t}(A|\bm{X}=\bm{x})} &= \frac{\partial_t \Cov_{P_t}(A,Y|\bm{X}=\bm{x})}{\Var_{P}(A|\bm{X}=\bm{x})} -  \frac{\Cov_{P}(A,Y|\bm{X}=\bm{x})}{\Var_{P}(A|\bm{X}=\bm{x})} \frac{\partial_t \Var_{P_t}(A,Y|\bm{X}=\bm{x})}{\Var_{P}(A|\bm{X}=\bm{x})}  \\
&= \frac{\tilde{f}(\bm{x})}{f(\bm{x})}\{\tilde{y}-\mu(\bm{x}) - \lambda(x)\{\tilde{a}-\pi(\bm{x})\}\}\frac{\tilde{a} - \pi(\bm{x})}{\Var(A|\bm{X}=\bm{x})}
\end{align*}
Thus, the result follow.

\section{Issues with the S-learner of the CATE}
\label{supp:slearn}

A simple alternative to the T-learner is the S-learner \citep{Kunzel2019}, with both being based on the decomposition $\tau(\bm{x})=\mu(1,\bm{x})-\mu(0,\bm{x})$. The S-learner estimate of the CATE is $\hat{\tau}^{(S)}(\bm{x})\equiv \hat{\mu}^*(1,\bm{x})-\hat{\mu}^*(0,\bm{x})$, where $\hat{\mu}^*(a,\bm{x})$ represents an estimate of $\mu(a,\bm{x})$ obtained by a regression of $Y$ on $(A,X)$ using all of the data. This is similar to the T-learner, except the S-learner uses a `single' regression $\hat{\mu}^*(a,\bm{x})$, and the T-learner uses `two' regressions to estimate $\hat{\mu}(1,\bm{x})$ and $\hat{\mu}(0,\bm{x})$. 

The main issue with the S-learner vs. the T-learner is that $\hat{\mu}^*(a,\bm{x})$ is chosen to make an optimal bias-variance trade-off over the population distribution of treatment and covariates. When there is poor overlap between the treated and untreated subgroups (e.g. treatment is correlated with covariates), then regularization biases, which control this trade-off, may bias the effect of treatment on outcome towards zero, making the S-learner potentially poorly targeted towards CATE estimation.

\begin{figure}[htbp]
\centering
\includegraphics[width=\linewidth]{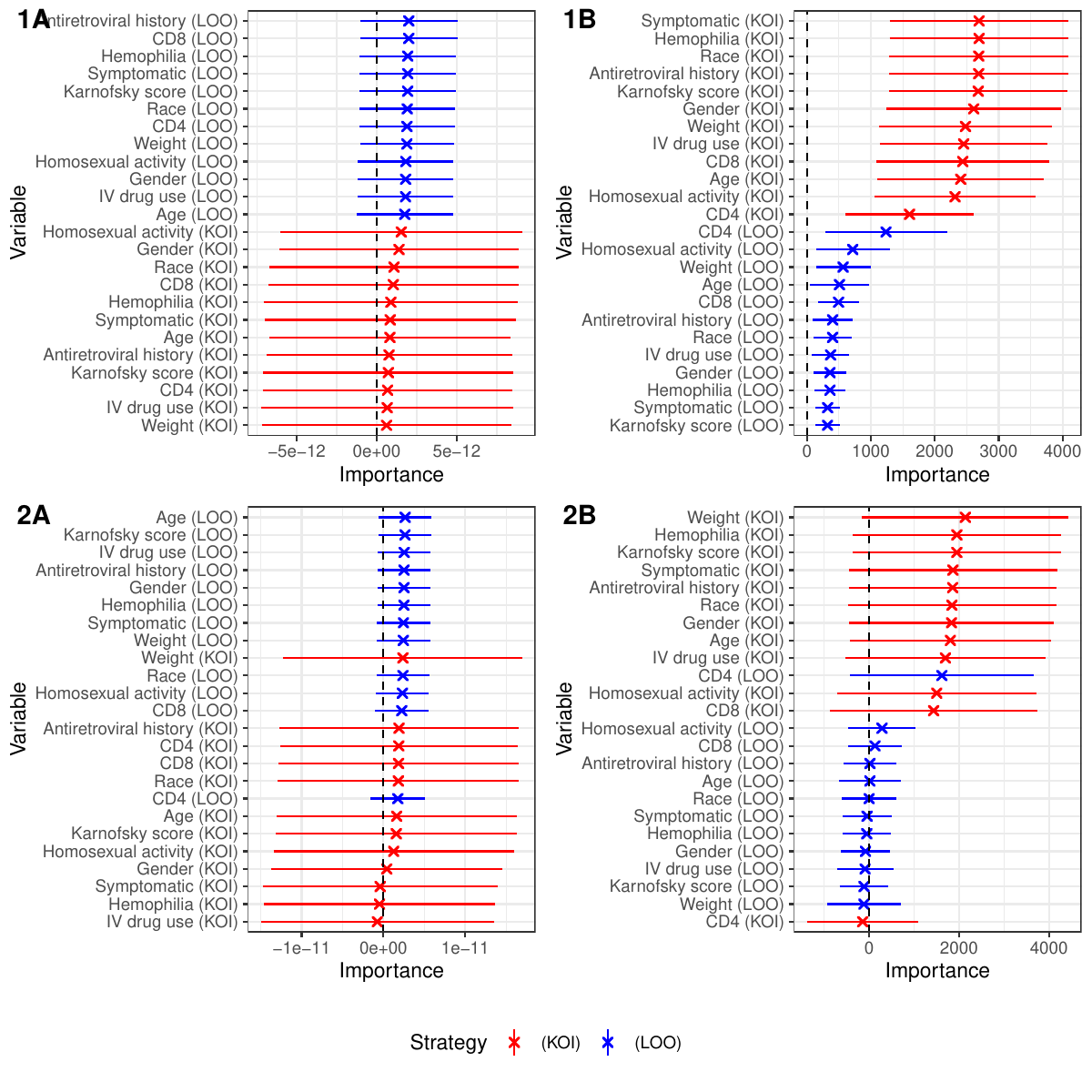}
\caption{TE-VIM estimates $\hat{\Theta}_s$ from the ACTG175 study using modified Algorithms. Error bars indicate 95\% CIs. In each plot, covariates are sorted according to their TE-VIM point estimate. Dashed lines indicate no importance. For the KOI mode, the TE-VIM represents the importance of the complement variable set, i.e. low-values denote high-importance of the KOI covariate.} 
\label{results_plot_unscaled_slearn}
\end{figure}

\begin{figure}[htbp]
\centering
\includegraphics[width=\linewidth]{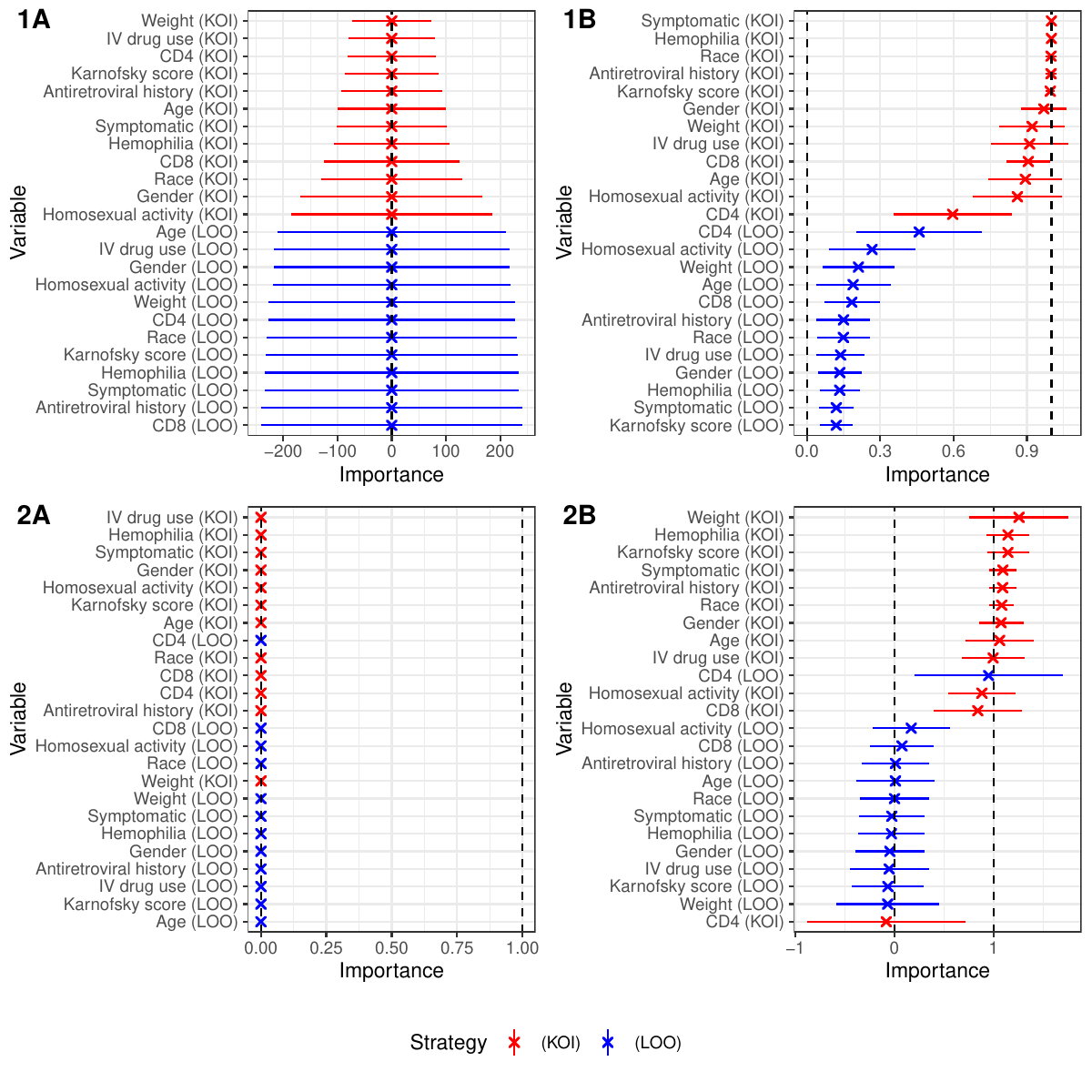}
\caption{Scaled TE-VIM estimates $\hat{\Psi}_s$ from the ACTG175 study using modified Algorithms. Error bars indicate 95\% CIs. In each plot, covariates are sorted according to their TE-VIM point estimate. Dashed lines indicate the $[0,1]$ support of the scaled TE-VIM. For the KOI mode, the TE-VIM represents the importance of the complement variable set, i.e. low-values denote high-importance of the KOI covariate.} 
\label{results_plot_scaled_slearn}
\end{figure}

In practice, this means that the S-learner is more likely to produce extremely small or even negative VTE estimates. For instance, if we replace $\hat{\mu}(a, \bm{x})$ with $\hat{\mu}^*(a, \bm{x})$ in all algorithms, then the results for our applied example are severely affected. In particular, Figures \ref{results_plot_unscaled_slearn} and \ref{results_plot_scaled_slearn} show the resulting unscaled and scaled TE-VIMs. We see that Algorithms based on the (modified) DR-learner (noSS-B and SS-B) are less affected when compared with the corresponding figures in the main paper, but Algorithms based on the S-learner (noSS-A and noSS-A) give wildly different results.

AIPW estimates of the ATE using the pseudo outcomes from the modified Algorithms noSS, SS-A, and SS-B were similar, respectively: 29.6$mm^{-3}$ (CI: 15.5, 43.8; p$<$0.01); 28.8$mm^{-3}$ (CI: 14.1, 43.5; p$<$0.01); 28.9$mm^{-3}$ (CI: 14.2, 43.7; p$<$0.01). VTE estimates differed substantially between S- and (modified) DR-learner based algorithms. With Algorithms noSS-A and SS-A returning negative point estimates: $-3.80\times10^{-9}mm^{-6}$ (CI: $-1.75\times10^{-9}$, $1.75\times10^{-9}$) and $-104mm^{-6}$ (CI: $-175$, $-33.0$), while Algorithms noSS-B and SS-B giving positive estimates: 2700$mm^{-6}$ (CI: 1300, 4090) and 1700$mm^{-6}$ (CI: -610, 4020). For Algorithms SS-A and SS-B we obtain the square root of these VTE estimates are 51.9 and 41.3$mm^{-3}$ respectively. Negative VTE estimates indicate that the S-learner of the CATE in Algorithms noSS-A and SS-A is a worse predictor of the pseudo-outcome than the sample mean pseudo-outcome (i.e. the AIPW estimate of the ATE based on $\hat{\mu}^*(a,\bm{x})$ and $\hat{\pi}(\bm{x})$).

\section{Applied example with Generalized Additive Models (GAMs)}
\label{supp:gam}

The applied analysis example using ACTG175 data was rerun with the `SuperLearner` regression steps replaced with Generalized Additive Model regression, as implemented via the `gam` package in R \cite{Hastie2004}. Figures \ref{results_plot_unscaled_slearn} and \ref{results_plot_scaled_slearn} show the resulting unscaled and scaled TE-VIMs. AIPW estimates of the ATE using the pseudo outcomes from the modified Algorithms noSS, SS-A, and SS-B were similar, respectively: 28.6$mm^{-3}$ (CI: 14.3, 43.0; p$<$0.01); 28.8$mm^{-3}$ (CI: 14.1, 43.4; p$<$0.01); 28.3$mm^{-3}$ (CI: 13.7, 43.0; p$<$0.01). VTE estimates from Algorithms noSS-A, noSS-B, SS-A, and SS-B were respectively: $2665mm^{-6}$ (CI: $1059$, $4271$; p$<$0.01), $2759mm^{-6}$ (CI: $1114$, $4405$; p$<$0.01), $1147mm^{-6}$ (CI: $-461$, $2756$; p$=0.16$), and $1255mm^{-6}$ (CI: $-468$, $2977$; p$=0.15$).

\begin{figure}[htbp]
\centering
\includegraphics[width=\linewidth]{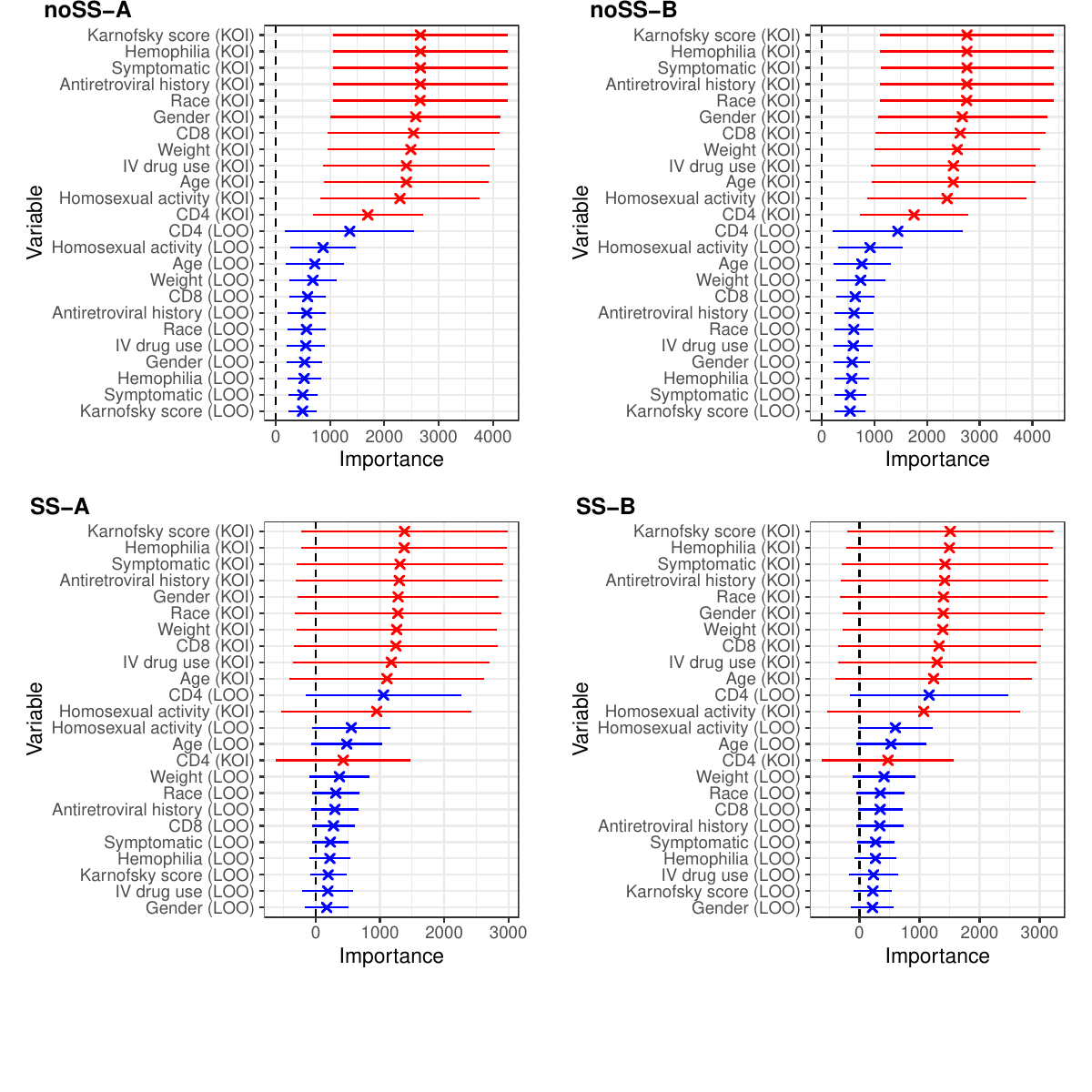}
\caption{TE-VIM estimates $\hat{\Theta}_s$ from the ACTG175 study using modified Algorithms. Error bars indicate 95\% CIs. In each plot, covariates are sorted according to their TE-VIM point estimate. Dashed lines indicate no importance. For the KOI mode, the TE-VIM represents the importance of the complement variable set, i.e. low-values denote high-importance of the KOI covariate.} 
\label{results_plot_unscaled_gam_tlearn}
\end{figure}

\begin{figure}[htbp]
\centering
\includegraphics[width=\linewidth]{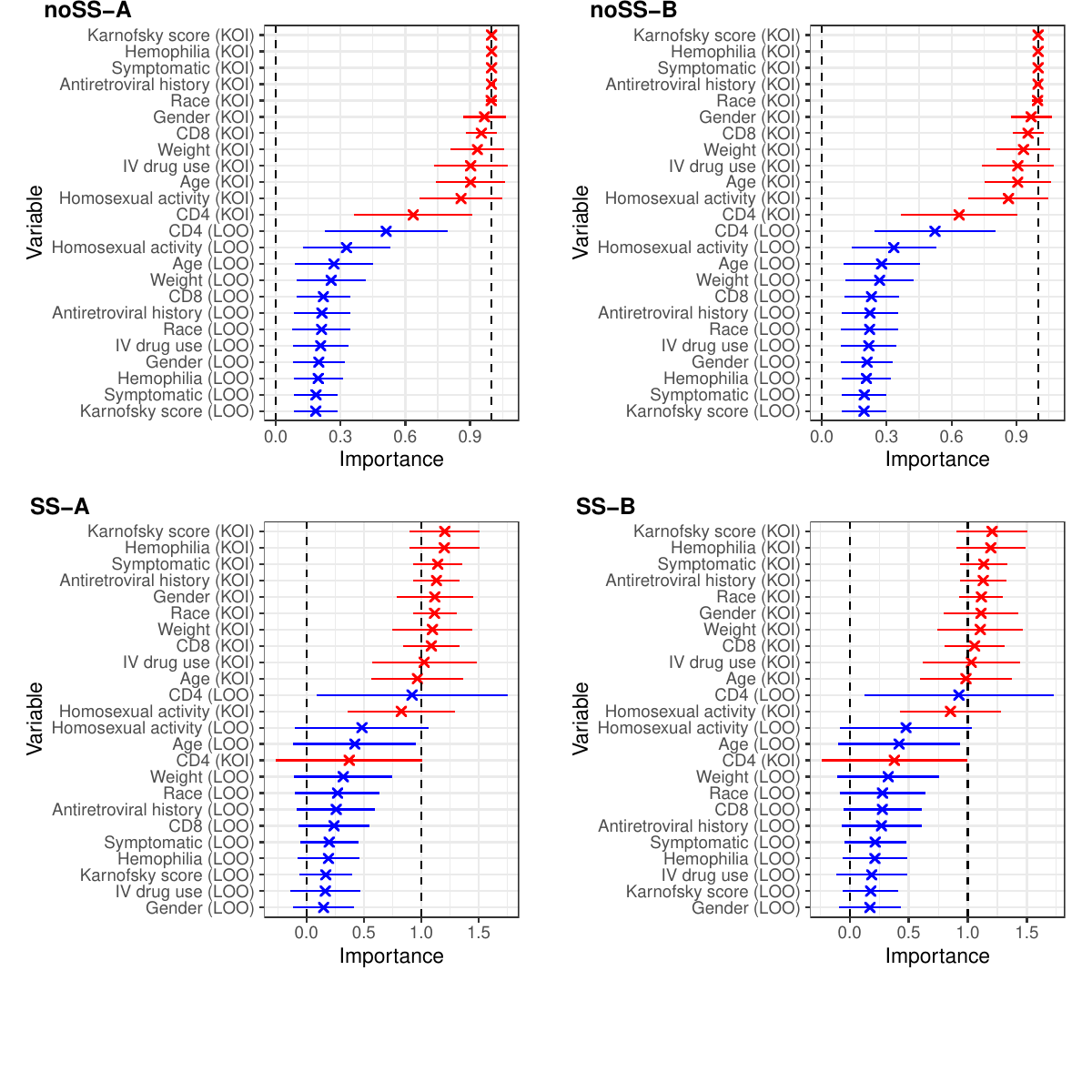}
\caption{Scaled TE-VIM estimates $\hat{\Psi}_s$ from the ACTG175 study using modified Algorithms. Error bars indicate 95\% CIs. In each plot, covariates are sorted according to their TE-VIM point estimate. Dashed lines indicate the $[0,1]$ support of the scaled TE-VIM. For the KOI mode, the TE-VIM represents the importance of the complement variable set, i.e. low-values denote high-importance of the KOI covariate.} 
\label{results_plot_scaled_gam_tlearn}
\end{figure}